\documentclass[10pt,english]{article}

\usepackage{amsmath, amsthm, amssymb}
\usepackage{a4, babel, graphicx, listings, fancyhdr, verbatim}
\usepackage{mathrsfs}
\usepackage{anysize}
\usepackage{natbib}
\usepackage[format = hang]{caption}

\usepackage{a4}
\usepackage[latin1]{inputenc}
\usepackage{amsmath, amsthm}
\usepackage{graphicx}
\usepackage{amssymb}
\usepackage{verbatim}
\usepackage{listings}
\usepackage{alltt}
\usepackage{babel}
\usepackage{mathrsfs}
\usepackage{ae}
\usepackage{natbib}
\usepackage{booktabs} 
\usepackage{color}
\usepackage{float}

\newtheoremstyle{theorem}
{10pt} 
{10pt} 
{\sl} 
{\parindent} 
{\bf} 
{. } 
{ } 
{} 

\theoremstyle{theorem}

\def\unif{\hbox{\rm unif}}

\def\cml{{\rm cml}}

\def\beq{\begin{eqnarray}}
\def\eeq{\end{eqnarray}}

\def\beqn{\begin{eqnarray*}}  
\def\eeqn{\end{eqnarray*}}

\def\E{{\rm E}}
\def\Var{{\rm Var}}

\def\dd{{\rm d}}
\def\N{{\rm N}}

\def\Pr{P}

\def\gold{{\rm gold}}
\def\eigen{{\rm eigen}}

\def\quadandquad{\quad \hbox{and} \quad}

\def\hatt{\widehat}
\def\tilda{\widetilde}

\def\eps{\varepsilon}
\def\half{\hbox{$1\over2$}}

\def\data{{\rm data}}
\def\alldata{{\rm all\ data}}
\def\obs{{\rm obs}}
\def\midd{\,|\,}
\def\tr{{\rm t}}
\def\dell{\partial}

\def\prof{{\rm prof}}

\def\con{{\rm conv}}
\def\fus{{\rm fus}}
\def\opt{{\rm opt}}

\def\Be{{\rm Be}}

\def\cc{{\rm cc}}

\def\ml{{\rm ml}}

\def\N{{\rm N}}
\def\binom{{\rm binom}}

\def\hatt{\widehat}
\def\tilda{\widetilde}
\def\sumjk{\sum_{j=1}^k}

\def\cc{{\rm cc}}
\def\obs{{\rm obs}}
\def\Gam{{\rm Gam}}
\def\cd{{\rm cd}}
\def\E{{\rm E}}
\def\half{\hbox{$1\over2$}}
\def\sgn{{\rm sgn}}

\def\prof{{\rm prof}}
\def\cprof{{\rm cprof}}
\def\eigen{{\rm eigen}}
\def\tr{{\rm t}}
\def\ml{{\rm ml}}
\def\midd{\,|\,}
\def\dd{{\rm d}}

\numberwithin{equation}{section}
\numberwithin{figure}{section}
\numberwithin{table}{section}
\usepackage{hyperref}

\title{Combining Information Across Diverse Sources: \\
   The II-CC-FF Paradigm}

\date{June 2020} 

\begin{document}


\maketitle

\centerline{\large\bf C\'eline Cunen and Nils Lid Hjort}


\medskip
\centerline{\bf $^1$Department of Mathematics, University of Oslo}

\begin{abstract}\noindent 
We introduce and develop 
a general paradigm for combining information across 
diverse data sources. In broad terms, 
suppose $\phi$ is a parameter of interest, built up via 
components $\psi_1,\ldots,\psi_k$ from data sources $1,\ldots,k$. 
The proposed scheme has three steps. First, the 
Independent Inspection (II) step amounts to investigating 
each separate data source, translating statistical 
information to a confidence distribution $C_j(\psi_j)$ 
for the relevant focus parameter $\psi_j$ associated 
with data source $j$. Second, Confidence Conversion (CC) 
techniques are used to translate the confidence distributions 
to confidence log-likelihood functions, say $\ell_{\con,j}(\psi_j)$. 
Finally, the Focused Fusion (FF) step uses relevant 
and context-driven techniques to construct a confidence 
distribution for the primary focus parameter 
$\phi=\phi(\psi_1,\ldots,\psi_k)$, acting on the combined 
confidence log-likelihood. In traditional setups, the II-CC-FF 
strategy amounts to versions of meta-analysis, and turns out to be competitive against state-of-the-art methods. Its potential lies in applications to harder problems, however. 
Illustrations are presented, related to actual applications.
\end{abstract}



\section{Combining information and the II-CC-FF scheme}
\label{section:intro}

Our paper concerns the statistical task of combining
information across different and perhaps very diverse
data sources. This is of course a long-standing 
theme in statistics, with papers going back to Karl Pearson
(cf.~\citet{SimpsonPearson04}); see 
\citet[Ch.~13]{CLP} for background and a general discussion 
of themes traditionally sorted under the bag-word meta-analysis, 
along with further basic references. The present paper aims 
at proposing and developing a certain paradigm, which we call 
the II-CC-FF method, meant to be powerfully applicable 
for ranges of situations far beyond the usual simpler
setups. We will explain the role and nature of 
the Independent Inspection (II), Confidence Conversion (CC), 
Focused Fusion (FF) steps below. 

A special case worth considering first is the textbook setup 
where $y_1,\ldots,y_k$ are independent estimators of
the same quantity $\psi$, and where 
$y_j\sim\N(\psi,\sigma_j^2)$, 
with known standard deviations $\sigma_j$. An easy
exercise in minimising variances shows that the 
optimally balanced overall estimator is 
\beq
\label{eq:simple}
\hatt\psi={\sumjk y_j/\sigma_j^2 \over \sumjk 1/\sigma_j^2}
   \sim\N\Bigl(\psi,\Bigl(\sumjk 1/\sigma_j^2\Bigr)^{-1}\Bigr).
\eeq 
A natural extension, though harder to analyse to full
satisfaction, is when $y_j\sim\N(\psi_j,\sigma_j^2)$,
with the individual means $\psi_j$ differing 
according to a $\N(\psi_0,\tau^2)$ distribution. 
For this type of random effects model, one wishes clear 
inference strategies for both the overall mean $\psi_0$
and level of variation $\tau$. We return to this
particular problem in 
Sections \ref{subsection:basicRE}, \ref{subsection:basicREsim}, 
and \ref{subsection:skulls}.

Many problems of modern statistics involving 
combining information are much more complicated than the 
situations 
sketched above, however. Sometimes one needs to 
combine `hard' data, with clear measurements from
controlled experiments, etc., with `soft' data, 
associated with information more loosely connected
to the parameters of primary interest, perhaps
via measurement errors or surrogate variables.
In addition there might be prior distributions
available, via subject matter experts, but
only for some of the parameters at play, not
enough to make it into a clear Bayesian analysis. 
For our development of II-CC-FF we have attempted 
to think fundamentally and generally about 
combination of information problems. Our framework 
encompasses known meta-analysis methods, but we aim at 
tackling new and more challenging problems as well. 
Parts of the meta-analysis literature are quite narrow, 
with specific methods for specific problems. 
In that light we hope our more general approach 
will be useful. 

\def\well{\diamond}
In reasonably general terms, assume there is 
a parameter $\phi$ of clear interest, related to 
parameters $\psi_1,\ldots,\psi_k$, either via 
a deterministic function $\phi=\phi(\psi_1,\ldots,\psi_k)$ 
or via some type of random effect distribution,
where such a $\phi$ might be a parameter 
related to a background distribution of the $\psi_j$.  
Suppose further that data source $y_j$ provides information 
pertaining to $\psi_j$. For the sake of clear presentation, 
we let the $\psi_j$ be one-dimensional here. 
Our II-CC-FF approach for reaching inference statements 
for the overall focus parameter $\phi$ can then be 
schematically set up as follows: 
 
\begin{itemize}
\item[$\well$] 
II, {\it Independent Inspection}: 
Data source $y_j$ is used, via appropriate models 
and analyses, to yield a confidence distribution $C_j(\psi_j,y_j)$
for the main interest parameter associated with study $j$. 
\item[$\well$] 
CC, {\it Confidence Conversion}: 
The confidence distribution is converted into a 
log-likelihood function for this main parameter of 
interest for study $j$, say $\ell_{\con,j}(\psi_j)$. 
\item[$\well$] 
FF, {\it Focused Fusion}: 
In the fixed effect case, the combined confidence log-likelihood 
function 
$\ell_\fus(\psi_1,\ldots,\allowbreak\psi_k)=\sumjk\ell_{\con,j}
(\psi_j)$ 
is used to reach focused fusion inference for 
$\phi=\phi(\psi_1,\ldots,\psi_k)$. With random effects, 
the fusion involves the computation of an integral.
\end{itemize} 

\noindent 
We do make use of certain subscripts 
in our paper, meant as helpful signposting. 
The subscript `$\con$' is for likelihood functions 
coming out the CC-step; `$\fus$' relates to the FF-step; 
while `$\prof$' and `$\cprof$' are used for the 
profile log-likelihood and its corrected version 
(see Section~\ref{subsection:corrections}). 

The extent to which some or all of these steps will
be relatively straightforward or rather complicated
to carry out depends to a high degree on the 
special features of the given source combination problem. 
The steps are not `isolated' or fully separated, 
but often related. 
In Section~\ref{subsection:automatic} we provide a standardised 
version of II-CC-FF, with a generic recipe to follow, 
but we will see that in many cases 
one could or should be more careful about the various steps.
In situations where the statistician
has all the raw data and the particular models used
for analysing the different sources of information,
the CC step is in a conceptual 
sense not difficult, as the required log-likelihood 
parts may be worked out from first principles. 
In various situations confronting the modern statistician 
this is rather more difficult, however, 
as one might have to base one's analysis 
on summary measures, directly or indirectly
given via other people's work, reports and publications. 
The II-CC-FF paradigm is meant to be powerfully 
applicable in such situations too. 

A pertinent question 
is whether or why there is a need for specific methods 
for combination of information in the first place;   
in a suitable sense, all of statistics concerns combination of 
information.  
One might therefore ask why there even exist subfields such as 
meta-analysis, and specific framework aimed at combination 
of information such as our own. So isn't meta-analysis just 
analysis? 
Two related responses are as follows. 
(i) Sometimes the full sets of data are not available,
with access only to summaries or partial summaries.
Issues here are storage, the practicalities of other
people's files, privacy concerns, etc. 
(ii) Sometimes it might be easier, conceptually or practically, 
to analyse the different sources or studies separately first, 
and then combine these pieces of summarised information.
Also, a statistical prediction is that modern statistics 
to an increasing degree will be concerned with 
such issues and challenges, finding and organising 
bits and pieces of information across different sources, 
with a need to reach conclusions based on these pieces. 

After a motivating illustration, below, 
we start in Section \ref{section:CDs} with a brief
review of confidence distributions (CDs), which are 
essential for the Independent Inspection (II) 
part of the programme. We then proceed with 
giving details related to the basics of 
Confidence Conversion (CC) in Section \ref{section:CC}
and Focused Fusion (FF) in Section \ref{section:FF}. 
In Section \ref{section:guides} 
we provide a standard version of our II-CC-FF framework, and 
investigate some pitfalls and solutions.
In Section \ref{section:meta} we investigate 
the use of our II-CC-FF scheme in well-established meta-analysis
situations, and in Section \ref{section:kina}
connections with other CD based approaches are explored.
Further performance and comparison issues are examined
in Section \ref{section:performance}, both via
simulations and decision theoretic risk functions.  
There, we find that II-CC-FF methods are competitive in several traditional
meta-analysis settings.
The three step II-CC-FF machinery is then seen in action
through four applications laid out in Section 
\ref{section:applications}.

\subsection*{Motivating illustration}
\label{section:motivation}


The following concrete illustration, which
has certain features placing it outside
the usual meta-analysis setups and methods,
shows the three steps of the II-CC-FF
at work, but with a minimum of details.  
The nonstandard aspect for this
illustration is partly that different studies
of the same statistical question have reported
different summary measures -- six
studies (call them type A) have reported
summary statistics based on continuous
outcomes, while five other studies
(which we call type B) reported
summaries based on a binary outcome.
More crucially, the focus parameter
$\beta$ in question, a regression coefficient 
related to the difference between treatments,
is not identifiable, and hence  
cannot be estimated directly,
for the type B studies. In fact, these
type B studies only inform us about
a certain  $\beta/\sigma$, where also
$\sigma$ is not identifiable, or estimable,
from those studies. Also, the raw data,
for these studies, are not available. 
The data employed here were first analysed  
in \citet{whitehead1999}; related problems 
have been treated in \citet{dominici2000} and \citet{LLX15}.

We have eleven randomised trials investigating the use of oxytocic 
drugs 
during labour and its potential effect on postpartum blood loss. 
Each study has two groups of patients, a treatment group receiving 
oxytocic drug and a control group receiving no drugs of that type.
Taking $y_{i,j}$ to be the blood loss for patient $i$ in study $j$, 
we may use the simple model $y_{i,j} = \alpha_j + \beta z_{i,j} + 
\eps_{i,j}$, 
with the $\eps_{i,j}$ independent and $\N(0,\sigma^2)$, 
and with $z_{i,j}$ an indicator variable, 
equal to 0 for patients in the control group and 
1 for patients in the treatment group. Here $\beta$ 
is the treatment effect and the parameter of main interest. 

For the six type A trials, we have 
the mean and the empirical standard deviation 
of the blood loss in the two groups of patients. 
With the simple normal model above, these four summary statistics 
are sufficient for each trial, and we thus have access 
to the full log-likelihood $\ell_{A,j}(\beta,\alpha_j,\sigma)$ 
for each continuous trial $j$. 
For the five type B trials, however, we merely have counts 
of the number of patients in each group having a blood loss 
of more or less than 500 ml. 
These numbers constitute a non-sufficient summary; 
we thus have less information in these studies 
compared to the continuous ones, and log-likelihood
functions not able to inform on $\beta$ directly,
only on $\beta/\sigma$. More specifically,  
based on the normal model above, we obtain a 
probit-type log-likelihood for these binary trials, say 
$\ell_{B,j}(\theta,\gamma_j)$, 
with $\gamma_j=(500-\alpha_j)/\sigma$ and $\theta=\beta/\sigma$.

Having made these modelling assumptions, the steps in the
II-CC-FF recipe follow straightforwardly. Using the
log-likelihood functions described above, we can, by methods
described in the next section, construct {\it confidence curves}
for the parameter of interest for each of the studies.

\begin{figure}[h] 
\begin{center}
\includegraphics[scale=0.5]{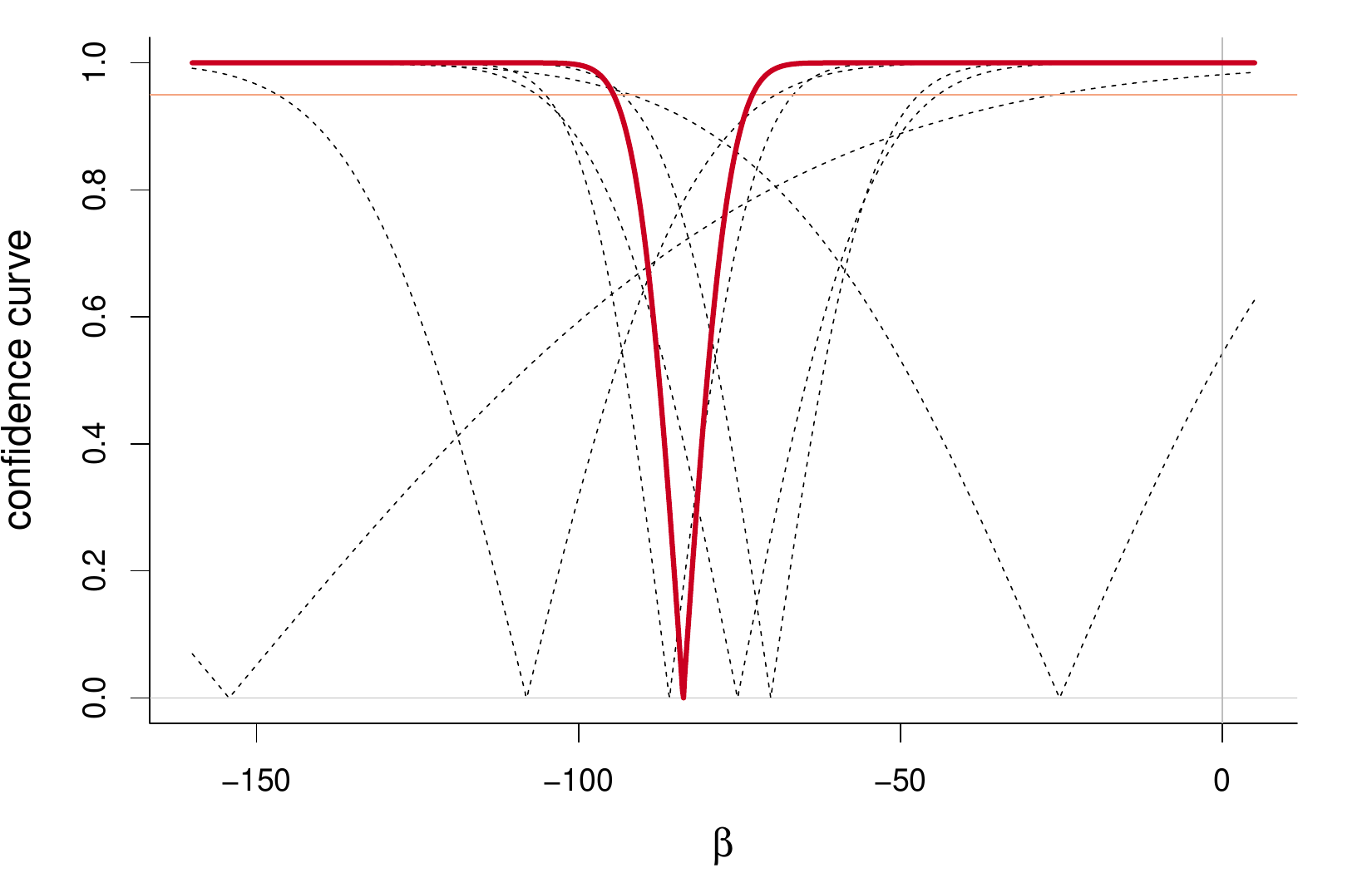}
\end{center}
\caption{Confidence curves for the treatment effect in the 
six continuous trials (dashed, black). In red, the confidence curve 
combining all the eleven studies.
The horizontal red line marks the 95\% confidence level.
The median confidence estimate is $-83.7$ ml, 
with 95\% interval $[-94.4, -73.1]$. 
\label{figure:blood} }
\end{figure}

The CC step is simple in this case, with no extra work
required, since the log-likelihood functions were used in the 
construction of the confidence curves for each study. 
In other situations we might have to carry out 
the conversion from confidence statements to log-likelihood
functions in ways described in Section \ref{section:CC}.
Via arguments explained in more detail in the next section, 
we reach log-likelihood contributions 
$\ell_{A,\prof,j}(\beta,\sigma)$ 
for the continuous studies and
$\ell_{B,\prof,j}(\beta/\sigma)$ 
for the binary studies.
In the FF step these are summed, to reach 
\beqn
\hbox{FF}\colon \qquad 
\ell_{\fus}(\beta,\sigma)
= \sum_{j=1}^6 \ell_{A,\prof,j}(\beta,\sigma) 
   +  \sum_{j=1}^5 \ell_{B,\prof,j}(\beta/\sigma). 
\eeqn 
Next we profile out $\sigma$ and obtain the final combined 
confidence curve by 
$$\cc^*(\beta,\alldata) 
   = \Gamma_1\bigl( 2\{ \max \ell_{\fus}(\beta,\hatt 
   \sigma(\beta)) 
   - \ell_{\fus}(\beta,\hatt \sigma(\beta)) \} \bigr), $$
with $\Gamma_1(\cdot)$ the c.d.f.~of a $\chi^2_1$.  
In Figure~\ref{figure:blood}, the thick red curve is this 
combined confidence curve. It is clearly narrower than all 
the individual curves and placed roughly in the middle of them, 
as we would expect. 

The combined inference clearly demonstrates that oxytocic drugs 
reduce postpartum blood loss, which is in agreement with 
the conclusions in \citet{whitehead1999}.
Here we have zoomed in on $\beta$ as the focus parameter,
to pinpoint precisely how much the two groups differ in blood loss.
For clinicians it might be of more direct interest 
to consider the probabilities for having a postpartum blood loss
greater than a threshold, like 500 ml, for the two groups, 
and then focus on the odds ratio, say $\rho$. Our approach can 
easily accommodate such an analysis too, with $\rho$ rather than
$\beta$ in the FF step, yielding a figure similar to 
Figure \ref{figure:blood}, but now for $\rho$.  
%
The II, CC, FF steps are not intended to form
a unique recipe, as there are individual variations,
depending on the application at hand. In the
FF step above the log-likelihood contributions
for type A and type B information were arrived at
via profiling of the fuller log-likelihoods constructed
under the auspices of the regression model 
we started out with. In other applications this 
would not be possible, and there would be a need
to convert CC information to log-likelihoods,
a theme we examine in Section~3.

\section{Independent Inspection: confidence distributions}
\label{section:CDs}

Suppose $Y_j$ denotes a set of random observations from data 
source $j$, stemming from a model with parameter $\theta_j$,
typically multidimensional, and with $\psi_j=\psi(\theta_j)$.
For the ease of presentation, we let $\psi_j$ be a one-dimensional 
focus parameter for now, but in general combination situations 
it will typically be multidimensional. 
A {\it confidence distribution} (CD) $C_j(\psi_j,y_j)$ for 
this focus parameter from source $j$ has the properties
(i) it is a cumulative distribution function 
(c.d.f.)~in $\psi_j$, for each $y_j$, and 
(ii) at the true value $\theta_0$, with associated 
true value $\psi_0=\psi(\theta_0)$, the distribution 
of $C_j(\psi_0,Y_j)$ is uniform on the unit interval. 
From this follows, under standard continuity and monotonicity
assumptions, that 
$${\Pr}_{\theta_0}\{C_j^{-1}(0.05,Y_j)
   \le\psi_0\le C_j^{-1}(0.95,Y_j)\}=0.90, $$ 
etc., i.e.~$[C_j^{-1}(0.05,y_{j,\obs}),C_j^{-1}(0.95,y_{j,\obs})]$ 
is a 90\% confidence interval for $\psi_j$, 
where $y_{j,\obs}$ denotes the observed dataset. Thus the CD
$C_j(\psi_j,y_{j,\obs})$, qua random c.d.f., is a compact and 
convenient
representation of confidence intervals at all levels,
and indeed a powerful inference summary. 
A close relative is the {\it confidence curve}, 
which we tend to prefer as a post-data graphical summary 
of information for focus parameters, defined as  
\beq
\label{eq:cc}
\cc_j(\psi_j,y_{j,\obs})=|1-2\,C_j(\psi_j,y_{j,\obs})|. 
\eeq 
It points to its cusp point, the median confidence
point estimate $\hatt\psi_{j,0.50}=C_j^{-1}(\half,y_{j,\obs})$, 
and the two roots of the equation $C(\psi_j,y_{j,\obs})=\alpha$ 
form a confidence interval with this confidence level. 
Degrees of asymmetry are easier to spot and to convey
using the confidence curve than with the cumulative
CD itself; cf.~illustrations
in Section \ref{section:applications}.  
We also note that the random $\cc_j(\psi_j,Y_j)$ 
has a uniform distribution, at the true position
in the parameter space, since $|1-2\,U|$ is uniform
when $U$ is. Indeed 
\beq
\label{eq:ccproperty}
{\Pr}_{\theta_0}\{\cc_j(\psi_0,Y_j)\le\alpha\}=\alpha, 
   \quad \hbox{for each\ }\alpha, 
\eeq
at the true parameters of the model. The confidence curve
is arguably a more fundamental concept than the confidence
distribution, as there are cases where a natural 
$\cc_j(\psi_j,Y_j)$ may be constructed, with a valid 
(\ref{eq:ccproperty}),
even when confidence regions are formed by disjoint intervals
(as with multimodal log-likelihood functions). 

For an extensive treatment of CDs,
their constructions in different types of setup, 
properties and uses, see \cite{CLP}, and the review paper
\cite{XieSingh13}, with ensuing discussion contributions.
The scope and broad applicability of CDs
are also demonstrated in a collection of papers 
published in the special issue {\it Inference With Confidence} 
of the journal {\it Journal of Statistical Planning and Inference},
2018 \citep{HjortSchweder18}. 
Here we shall merely point to two important
and broadly useful ways of constructing a confidence
distribution, for a focus parameter $\psi_j$, based 
on data from a model with a multidimensional parameter 
$\theta_j$. 
The first is to rely on an approximately normally 
distributed estimator, if available, say 
$\hatt\psi_j\sim\N(\psi_j,\kappa_j^2)$, and with 
standard deviation
well estimated with an appropriate $\hatt\kappa_j$. Then,
with $\Phi(\cdot)$ as usual denoting the c.d.f.~of the standard 
normal,  
$C_j(\psi_j,y_j)=\Phi((\psi_j-\hatt\psi_j)/\hatt\kappa_j)$
is an approximately correct CD, 
first-order large-sample correct under weak regularity
conditions. In particular the estimator used can be
the maximum likelihood one (ML), say $\hatt\psi_{j,\ml}$, 
but other estimators are allowed too in this simple 
construction. The second is based on the profiled 
log-likelihood function 
$\ell_{\prof,j}(\psi_j)=\max\{\ell_j(\theta_j)\colon \psi(\theta_j)=\psi_j\}$,
which leads to the deviance function 
\beq 
D_j(\psi_j)=2\{\ell_{\prof,j}(\hatt\psi_{j,\ml})-\ell_{\prof,j}(\psi_j)\}
   =2\{\ell_{\prof,j,\max}-\ell_{\prof,j}(\psi_j)\}. 
\label{eq:deviance}
\eeq
As laid out in \citet[Chs.~2, 3]{CLP}, 
the Wilks theorem with variations then lead naturally to 
\beq
\cc_j(\psi_j,y_j)=\Gamma_1(D_j(\psi_j)),
\label{eq:ccdeviance}
\eeq 
with $\Gamma_\nu(\cdot)$ denoting the c.d.f.~of a 
$\chi^2$ with degrees of freedom $\nu$.
Typically, the second method (\ref{eq:ccdeviance}) leads
to a better calibrated confidence curve than the 
the simpler method mentioned first. 

\section{Confidence Conversion: from confidence to likelihoods} 
\label{section:CC}

Several well-explored methods, with appropriate variations 
and amendments, lead from likelihood functions to 
CDs and confidence curves; 
cf.~again several chapters of \citet{CLP}. Sometimes
the CC step comes almost for free, in cases where 
the statistician can compute say log-likelihood 
profiles from raw data, or from sufficient statistics, 
for the given models. 
But in some cases the  CC step of the II-CC-FF paradigm 
requires methods for going the other way, from CDs
or confidence curves to log-likelihood information, 
and this is more involved.  Among the complications 
is that different experimental protocols, 
with ensuing different CDs,
might be having the same log-likelihood functions, 
so the link between confidence and likelihood is not 
one-to-one. 

\citet[Ch.~10]{CLP} develop and discuss this topic 
at some length. For the present purposes we shall be
content with what we call the {\it chi-squared inversion},
associated with (\ref{eq:ccdeviance}) above. 
Assume that all our information about a parameter $\psi_j$ from 
source $j$ comes in the form of the confidence curve  
$\cc_j(\psi_j,y_j)$. Then we can obtain a confidence log-likelihood 
contribution from source $j$ by the following formula 
\beq
\ell_{\con,j}(\psi_j)=-\half\Gamma_1^{-1}(\cc_j(\psi_j,y_j)).
\label{eq:ccinversion}
\eeq   
If the confidence curve has been constructed via the 
second general method presented in the previous section,
see (\ref{eq:ccdeviance}), we will of course simply get back the 
profiled log-likelihood function. The CC step therefore only comes 
into play when confidence curve are constructed via non-standard 
methods, as we will see in
Applications \ref{subsection:whales} and \ref{subsection:quantiles}, 
or when the only available information from source $j$
is the confidence curve itself and we do not know exactly
how it was constructed. 

When a CD is available, rather than a confidence curve,
one can use the {\it normal conversion} 
$\ell_{\con,j}(\psi_j)=-\half\{\Phi^{-1}(C_j(\psi_j,y_j))\}^2$.
This is equivalent to the recipe in \eqref{eq:ccinversion}, 
when the confidence curve has been constructed via 
$\cc_j(\psi_j,y_j)=|1-2\,C_j(\psi_j,y_j)|$.
A relevant point here is that one often constructs
a confidence curve $\cc_j(\psi_j,y_j)$ directly, not 
always via (\ref{eq:cc}), making (\ref{eq:ccinversion})
a more versatile tool. 
The normal conversion confidence likelihood is also
what \citet{Efron93} proposed, for coming from confidence
to likelihood, via different arguments and for different purposes;  
see also \citet[Ch.~11]{EfronHastie16}. 
For details on how well the chi-squared inversion
methods works, in different scenarios, see
Section \ref{subsection:fusionlemmas}. 

In some situations one is able to  construct a CD for source $j$
via a one-dimensional statistics $T_j$, instead of using the
general method from \eqref{eq:ccdeviance}. Then one may use
{\it  exact conversion} to obtain the confidence log-likelihood.
When the statistic has a continuous distribution,
the exact conversion of the CD $C_j(\psi_j,T_j)$,
see \citet[Ch.~10]{CLP}, is
$\ell_{\con,j}(\psi_j)=\log\,\lvert \dell C_j(\psi_j,t)/ \dell t\rvert$.

\section{Focused Fusion: from full likelihood
     to focus parameter}
\label{section:FF}

Suppose now that the II and CC steps have been 
successfully carried out, leading to confidence log-likelihood 
contributions $\ell_{\con,j}(\psi_j)$ from information sources
$j=1,\ldots,k$. Depending on the application and its context
we might then be interested in either a fixed effect approach,
with the main focus parameter $\phi$ is a function 
of the $\psi_j$, or a random effect approach, where we introduce 
an additional layer of heterogeneity through a model 
for the $\psi_j$. We will treat the fixed effect case first.   

\subsection{Fixed effects fusion}

Assuming the information sources to be independent,
the overall confidence log-likelihood function is 
$\ell_\fus(\psi_1,\ldots,\psi_k)=\sumjk \ell_{\con,j}(\psi_j)$. 
When focused inference is wished for, for a focus parameter 
$\phi=\phi(\psi_1,\ldots,\psi_k)$, the natural way forward
is, again, via profiling: 
\beqn
\ell_{\fus,\prof}(\phi)=\max\{\ell_\fus(\psi_1,\ldots,\psi_k)
   \colon\phi(\psi_1,\ldots,\psi_k)=\phi\}. 
\eeqn  
By the Wilks theorem directly, or by variations of the
arguments and details used to prove such theorems 
(cf.~\citet[Appendix]{CLP}), the overall deviance function 
\beqn
D^*(\phi)
   =2\{\ell_{\fus,\prof}(\hatt\phi)-\ell_{\fus,\prof}(\phi)\} 
\eeqn 
tends, at the true parameter position and with 
increasing information volume, to a $\chi^2_1$. Here 
$\hatt\phi$ is the ML, maximising the profiled log-likelihood.
Hence 
\beq
\label{eq:ccstar}
\cc^*(\phi,\alldata)=\Gamma_1(D^*(\phi)) 
\eeq 
is the outcome of the three step II-CC-FF machine, 
a confidence curve for the focus parameter. 
In Section \ref{subsection:fusionlemmas} we will come back 
to some discussion on the meaning of 
`increasing information volume' in a combination context. 
In situations where the $\psi_j$ represent the 
same focus parameter, common across sources, the 
scheme above simplifies. 

\subsection{Random effects fusion}
 \label{subsection:random}
 
In our II-CC-FF setting, we use the term `random effects' 
when we wish to introduce an extra layer of heterogeneity 
in the fusion step. This is more easily presented when assuming 
that $\psi_1, \dots, \psi_k$ are scalars. In the random effects 
case we do not assume that all $\psi_j$ are equal but rather 
that they come from some underlying distribution.
In the most canonical case, this distribution 
will be governed by some overall mean parameter $\psi_0$ 
and some spread parameter $\tau$; specifically we could have 
$\psi_j\sim\N(\psi_0,\tau^2)$. The parameter of main interest 
may be either the overall mean, or the spread, 
or perhaps a quantile, depending on the context. 

We propose the following general solution for II-CC-FF 
with random effects. Suppose the $\psi_j$ are modelled 
as coming from a background density $f(\psi_j,\kappa)$,
say, where the $\kappa$ could be a centre and a spread 
parameter, as for $(\psi_0,\tau)$ in the normal case. 
Then, using the confidence log-likelihoods 
$\ell_{\con,j}(\psi_j)$ from each source, we define the fusion log-likelihood to be 
\beq
\ell_{\fus}(\kappa) 
   = \sumjk \log\Bigl[\int \exp\{\ell_{\con,j}(\psi_j)\}f(\psi_j,\kappa)
   \,\dd\psi_j\Bigr].
\label{eq:ffrandom}
\eeq 
We would usually need to profile again, 
depending on what we are interested in, 
say the centre $\psi_0$ or spread $\tau$ for the 
case of a normal model for the $\psi_j$. 
To produce our final confidence curve we will often use the 
Wilks approximation. 
This II-CC-FF solution requires the computation of integrals. 
Sometimes numerical integration routines in R work well enough, 
other times we will make use of the so-called Template Model 
Builder package (TMB) and its Laplace approximations in order 
to compute the integral \citep{Kristensen2016}.

\subsection{Wilks theorems for conversion and fusion}
\label{subsection:fusionlemmas}

There are chi-squared approximation methods at work
at sometimes several levels in our II-CC-FF scheme.
For some applications the chi-squared inversion
method (\ref{eq:ccinversion}) is crucial,
as for the nonparametric CDs for quantiles
in Application \ref{subsection:quantiles},
and for other situations what matters more might
be the chi-squared approximation of the FF step (\ref{eq:ccstar}).  
Limit distribution results securing such
$\chi^2_1$ limits are collectively referred to as
Wilks theorems, with different setups of regulatity
conditions. We refrain from setting up lists of
precise regularity conditions here, as applications
of the theory would involve different types
of situations, but we give brief pointers
to relevant methods and literature, as follows.

First, regarding (\ref{eq:ccinversion}) and
conversion to log-likelihoods,
both the exact log-likelihood and the inversion
approximation are guaranteed to be close to the negative quadratic 
$-\half(\psi_j-\hatt\psi_{j,\ml})^2/\hatt\kappa_j^2$, 
for the appropriate $\hatt\kappa_j$, by arguments
associated with classical large-sample calculus.
This would include asymptotic normality of the ML
estimator and indeed the traditional Wilks theorem,
see \citet[Ch.~2 and Appendix]{CLP}. 
The resulting approximations are typically good
also when the data information volume is small,
as long as the underlying models are smooth
in their parameters.
Second, the arguments and methods pointed to also
entail that the FF inference method (\ref{eq:ccstar})
is large-sample close to that of minimising the relevant 
$\sumjk(\psi_j-\hatt\psi_{j,\ml})^2/\hatt\kappa_j^2$
under $\phi=\phi(\psi_1,\ldots,\psi_k)$
constraints. For such minimum chi-squared methods, 
precise Wilks theorems are given in
\citet[Section 23]{Ferguson96}.
Regularity conditions there are of the type
where the number of information sources $k$
is kept moderate and fixed, but with steadily
more data for each. 
Importantly, limiting normality of estimators,
along with limiting $\chi^2_1$ results for deviances,
can also be derived in the rather different setups
with small data volume for each source, but
where the number $k$ increases. A case in point
is where $\phi=b^\tr\psi$ is linear in the $\psi_j$,
with estimator $\hatt\phi=b^\tr\hatt\psi$,
and the FF step is large-sample equivalent to
$D^*(\phi)=(\phi-\hatt\phi)^2/\sumjk b_j^2\hatt\kappa_j^2$. 
This tends to a $\chi^2_1$ for increasing $k$,
under mild regularity conditions. 

\section{II-CC-FF versions}
\label{section:guides}

The overall objective of the II-CC-FF is to construct 
a valid confidence curve for each parameter $\phi$ 
of particular interest, typically of the form 
$\phi=\phi(\psi_1,\ldots,\psi_k)$, incorporating the relevant 
information in all the sources. 
The framework we have presented so far has not intended
to provide a single clear-cut recipe for doing such analyses
in practice. 
Below one such concrete recipe is presented, however, 
which we call the standard II-CC-FF method, and which
may be used for a wide range of models and data.
The standard framework has limitations, which we will 
discuss, and which will serve as a starting point for the 
presentation of some partial solutions, and more fine-tuned 
versions of the II-CC-FF scheme.
These discussions also highlight various important general 
issues with methods for combination of information which 
are relevant also outside the II-CC-FF framework.

\subsection{Standard II-CC-FF}
\label{subsection:automatic}

The scheme to be described now requires that
we have the  full data available, or sufficient summaries,  
from all sources. The statistical work starts by 
deciding on one or more parameters of particular interest,
involving relevant parameters $\psi_1,\ldots,\psi_k$
from the $k$ sources. These might be parameter vectors
(i.e.~need not be one-dimensional),
they might differ from source to source,
but may also contain common parameters across sources. 

\begin{enumerate}
\item[$\well$] II, {\it Independent Inspection}: 
analyse each source $j$ separately. Assume a parametric model for 
the observations and put up the likelihood function. 
Profile out the source-specific nuisance parameters, and obtain 
$\ell_{\prof,j}(\psi_j)$.
\item[$\well$] CC, {\it Confidence Conversion}: 
in this case we already have the log-likelihood profiles 
from each source, so the confidence conversion is simple,  using the $\ell_{\prof,j}(\psi_j)$ directly.
\item[$\well$] FF, {\it Focused Fusion}: 
here we want to obtain a confidence curve for the parameter 
of overall interest $\phi$. Depending on the situation,
(i) if $\phi$ is assumed to be the same across sources or a function 
of some source-specific parameters, sum the $\ell_{\con,j}$ 
and then profile again if necessary;
(ii) if some component of the $\psi_j$ are assumed 
to come from some common distribution, use the random effects 
solution presented above, and then profile again if needed.
We then obtain $\ell_{\fus,\prof}(\phi)$, and in both cases 
we use the Wilks approximation to produce the final, 
combined confidence curve $\cc^{*}(\phi,\data)$. 
\end{enumerate}

As for confidence curves in general, we consider the method 
to work if the final combined confidence curve $\cc^{*}(\phi)$ 
has the right coverage properties, 
either exactly or approximately, as per (\ref{eq:ccproperty}). 
If the final combined confidence curve does not have the
correct coverage properties, this may be due to two related 
problems:
(1) the profiling in either the II or the FF step has gone wrong; and 
(2) the distribution of the deviance (based on the profile
log-likelihood) is far from a $\chi^2_1$,
i.e.~the Wilks approximation is not valid. 

Problem (2) is related to the issues discussed in Section \ref{subsection:fusionlemmas} 
and  will usually disappear when either $k$ or the $n_j$ increase.
In situations with little data, the Wilks approximation
can sometimes be ameliorated using relatively simple tools,
like the Bartlett correction; 
see for instance \citet[Chs.~7, 8]{CLP} for a discussion 
on such fine-tuning methods and second-order approximations. 
Further, in some situations one may be able to derive and simulate 
the distribution of the deviance exactly, and thus bypass the use of 
the Wilks approximation altogether. We will see examples of such 
II-CC-FF versions in Sections \ref{subsection:basicRE} and 
\ref{subsection:skulls}.

Problem (1) is related to the profiling and the presence of nuisance 
parameters. In situations with nuisance parameters using 
the profile log-likelihood can lead to 
``inefficient and even inconsistent estimates'' 
\citep{MccullaghTibshirani1990}.
 As we see above, the standard II-CC-FF method may often require 
two rounds of profiling: 
first in the II step where we might profile out the 
\textit{source-specific} nuisance parameters, and sometimes 
in the FF step where we might profile out \textit{shared} nuisance 
parameters (which are shared by the $k$ sources). If we have 
`large sources', i.e.~the sample size $n_j$ 
of each source is large, 
we can safely profile in the II step. 
If some or all the sources are small, however, one 
should be more careful. Specifically, the profiling 
might go wrong and we illustrate this situation 
with a famous example in Appendix \ref{section:appendixA}, the
Neyman-Scott problem.

Shared nuisance  parameters can be of different kinds, 
and here we will particularly concern ourselves with 
nuisance parameters arising from the random effect distribution 
in the FF step. For example, if we have 
$\psi_j\sim\N(\psi_0,\tau^2)$ and our focus parameter 
is $\psi_0$, then $\tau$ is a shared nuisance parameter of that 
type. 
If the number of sources $k$ is large we can safely profile, while if 
$k$ is small we may 
need to resort to some of the corrections described next.
Often we  have both source-specific
and shared nuisance parameters. 
 In that case, 
we ideally need to have a large number of large 
sources in order to produce  valid CDs 
with the default profiling-based method. 
Note that in these cases,  if $k$ is too small, 
large sources will not necessarily help. Conversely, 
if the $n_j$ are too small, a large $k$ will not 
in general be able to remedy the mistakes coming 
from profiling in the II step.

\subsection{Corrections to the log-likelihood profile}
\label{subsection:corrections}

There is a large literature
concerning corrections or 
modifications to the profile likelihood.
The  different corrections appearing in the literature 
have varying performance and complexity; 
see for instance \citet{Barndorff1986},
\citet{CoxReid1993}, \citet{DiciccioEfron1992}, 
\citet{Stern1997}, \citet{Diciccio1996}. 
There is also a whole subfield of integrated likelihood methods 
with partly similar aims, see \citet*{Berger1999}. 
A thorough investigation of all these methods 
is outside the scope of this article, and we will 
therefore only present one rather simple, somewhat 
limited solution. Alternative methods might work better, 
or at least in a more general setting, but these are often 
more complicated to compute.

In \citet{CoxReid1987}, the authors present what we 
will term the simple Cox--Reid correction. This is possibly 
the easiest correction to compute among those suggested above. 
It can be considered a special case of the correction 
in the general modified profile likelihood of Barndorff-Nielsen, 
but the simple Cox--Reid correction is limited 
to situations with orthogonal parameters (i.e.~that 
the off-diagonal terms in the expected information matrix 
are equal to zero).
Assume we have a scalar parameter of interest $\psi$ 
and some vector of nuisance parameters $\lambda$. As usual, 
the profile log-likelihood for $\psi$ is defined as 
$\ell_{\prof}(\psi) = \ell(\psi,\hatt\lambda(\psi))$,
where $\hatt\lambda(\psi)$ is the ML estimate of $\lambda$ 
for each fixed $\psi$ value. The simple Cox--Reid correction 
gives the following modification of the profile log-likelihood, 
\beq
\label{eq:corr_prof}
\ell_{\cprof}(\psi) = \ell_{\prof}(\psi) 
   - \half \log[\det J_{\lambda \lambda}(\psi, \hatt\lambda(\psi)) ]
\eeq 
where 
$J_{\lambda \lambda}(\psi, \hatt \lambda(\psi))
=- \dell^2 \ell(\psi,\lambda) / (\dell \lambda \dell \lambda^\tr)$
is the observed information for the $\lambda$ components,
evaluated at $(\psi, \hatt \lambda(\psi))$.
The simple Cox--Reid correction can be used both in the II and FF steps,  
for models with orthogonal parameters. 
We illustrate the use of the Cox--Reid correction in the II step in 
the 
Appendix Section \ref{section:appendixA}.
In the FF step, corrections may be necessary when there are shared 
nuisance parameters arising from a random effect distribution. In 
particular, we propose that this correction should readily be applied 
when then random effect distribution is assumed to be normal. 
Here, the correction should be particularly notable for small $k$. 
We will present some Cox--Reid corrections in a classic model for 
random effect meta-analysis in Section \ref{section:meta}.

\subsection{Optimal CD methods} 
\label{subsection:torenilsmethod} 

For some parameters in exponential families, we can bypass the standard II-CC-FF, and 
its potential problems, by making use of the following 
alternative method which is much more powerful, 
producing \textit{optimal CDs}. 
In our II-CC-FF setting, this method might come into play 
both in the II step and the FF step, 
see the application in Section \ref{subsection:meta2x2}.

For ease of presentation, we present the optimal confidence 
method in the case where all the $k$ sources inform on 
a common focus parameter $\psi=\psi_1=\dots=\psi_k$. This constitutes 
a situation where the method is used in the final FF step.
Suppose again that $\psi$ is the focus parameter, and that 
we have $m$ nuisance parameters $\gamma_1,\dots,\gamma_m$, 
which may be both source-specific or shared across all $k$ 
sources. 
Suppose also that the log-likelihood function at work, 
based on information sources $y_1,\ldots,y_k$,  
can be written in the form 
\beq
\label{eq:optimal}
\ell(\psi,\gamma_1,\ldots,\gamma_m)
   =\psi A+\gamma_1B_1+\cdots+\gamma_mB_m - d(\psi,
   \gamma_1,\ldots,\gamma_m)
   +h(y_1,\ldots,y_k), 
\eeq 
where $A$ and $B_1,\ldots,B_m$ are statistics, 
i.e.~functions of the data collection, with 
observed values $A_\obs$ and $B_{1,\obs},\ldots,B_{m,\obs}$, 
and with $m$ often bigger than $k$. 
Then, under mild regularity conditions, 
there is an overall most powerful CD,
namely 
\beqn
C^*(\psi,y)={\Pr}_{\psi}\{A\ge A_\obs\midd 
   B_1=B_{1,\obs},\ldots,B_m=B_{m,\obs}\}. 
\eeqn 
That this $C^*(\psi,y)$ indeed depends on $\psi$ 
but not on the $\gamma_j$ parameters is part of the 
result and the construction. 

To illuminate the exact meaning of `most powerful' 
in this setting, one needs to consider the theory for loss 
and risk functions for CDs developed in \citet[Ch.~5]{CLP}.
Confidence power is measured via the risk function 
\beq
\label{eq:risk}
r(C,\psi,\gamma)=\E_{\psi,\gamma}\int \Gamma(\psi_\cd-\psi)\,
\dd C(\psi_\cd,Y), 
\eeq 
for any convex nonnegative $\Gamma(\cdot)$ with $
\Gamma(0)=0$.
The random mechanism involved in the expectation here
is a two-stage operation; first data $y$, governed by 
the $(\psi,\gamma)$ held fixed, are used to generate
the CD $C(\psi,y)$, and then  
$\psi_\cd$ is a random draw from this distribution. 
Intuitively, a low confidence risk means that 
the CD in question is tight around true value of $\psi$, 
while CDs which are less concentrated around the true value 
will have a higher risk. A CD with 
low confidence risk should therefore be expected to produce 
narrow confidence intervals (but keeping the correct coverage), 
and point estimates with little bias. We will see the 
confidence risk concept at work in Section \ref{subsection:nils}.

\section{Meta-analysis}
\label{section:meta}
   
 
As  mentioned in the small start example \eqref{eq:simple}, 
some common meta-analysis methods flow more or less directly 
from the II-CC-FF framework. In addition, the framework 
also invites more general, principled and non-standard 
solutions some which we will explore in this section. 
Further, we will investigate connections 
between II-CC-FF and a couple of widely encountered 
meta-analysis methods. We will start with a discussion 
of the basic random effect model, before we go on to the famous 
case of meta-analysis of $2 \times 2$ tables. 

\subsection{The basic random effect model}
\label{subsection:basicRE}
 
The most canonical type of random effect meta-analysis, 
which we term the basic random effect model, starts with $k$ 
independent estimators $y_1, \dots, y_k$ aiming at 
the parameters $\psi_1, \dots, \psi_k$, 
with $y_j \midd \psi_j \sim \N(\psi_j, \sigma_j^2)$, 
and $\psi_j \sim \N(\psi_0, \tau^2)$. Usually the source-specific 
standard deviations $\sigma_j$ are assumed known. The literature 
treating this model is enormous, see for instance 
\citet{langan2019} and \citet{PartlettRiley2017} 
and references therein. Note that likelihood-based method for the 
basic random effect model, 
even exploring higher order corrections, have been 
investigated earlier, see for instance \citet{Hardy1996} 
and \citet{Noma2011}. For a more general likelihood approach 
see \citet{ORourke2008}.

When assuming known $\sigma_j$, the integral in 
\eqref{eq:ffrandom} has an explicit solution and the standard 
II-CC-FF solution will rely on the following log-likelihood function 
\beqn
\ell(\psi_0, \tau) 
= \sumjk \Bigl\{ -\half \log(\sigma_j^2+\tau^2) 
   - \half \frac{(y_j - \psi_0)^2}
   {\sigma_j^2+\tau^2}  \Bigr\},
\eeqn
where we profile out either $\psi_0$ or $\tau$ depending on which 
parameter is of main interest. The confidence curves for $\psi_0$ 
and $\tau$ will point at the standard ML estimators,
and the ensuing confidence intervals will be very similar 
to solutions which have been investigated in some of the references 
mentioned above. These solution are reasonably good when $k$ is 
not too small, see also Section \ref{subsection:basicREsim}, but 
may be improved upon. \citet{langan2019} finds, for instance, 
that the ML estimator for $\tau$ has relatively poor performance in 
terms of bias and mean squared error. 

We can attempt to improve on the standard II-CC-FF solution using 
the Cox--Reid correction. First we will consider the case where 
$\psi_0$ is the parameter of main interest, then
the full combined profile likelihood from the FF step becomes 
\beq
\begin{array}{rcl}
\ell_{\fus,c\prof}(\psi_0) 
  &=&\displaystyle
      \sumjk \Bigl\{ -\half \log(\sigma_j^2+\hatt \tau^2(\psi_0)) 
   - \half \frac{(y_j - \psi_0)^2}
   {\sigma_j^2+\hatt \tau^2(\psi_0)}  \Bigr\} \\
  & &\displaystyle
      \qquad\qquad 
   - \half  \log \sumjk \Bigl\{ 
   - \half \frac{1}{(\sigma_j^2+\hatt \tau^2(\psi_0))^2} 
   + \frac{(y_j - \psi_0)^2}{(\sigma_j^2+\hatt \tau^2(\psi_0))^3}  
    \Bigr\}.
\end{array}
\label{eq:ffREpsi}
\eeq 
The first part of the formula is the ordinary profile log-likelihood, 
the second part the simple Cox--Reid correction.  We obtain the 
combined confidence curve using the Wilks approximation. This 
correction has been obtained by differentiating with respect to 
$\tau^2$ in \eqref{eq:corr_prof}; note that a somewhat 
different correction term would have been obtained
if we had differentiated with respect to $\tau$.
We have not seen this solution anywhere in the literature, and we 
investigate its performance in Section \ref{subsection:basicREsim}. 
In this situation $\tau$ is a `border parameter',
as the profiled ML estimate $\hatt\tau(\psi_0)$ can be zero
with positive probability. This happens precisely when 
\beq
\label{eq:condTau}
\sumjk \frac{1}{\sigma_j^2 }\Bigl\{ \frac{(y_j - \psi_0)^2}
   {\sigma_j^2} -1  \Bigr\} \le 0, 
\eeq 
and if this takes place there is an interval of
such $\psi_0$ values. The correction term will then
experience non-smoothness at the end points of that interval. 
To avoid this non-smoothness issue, where the
rationale behind the Cox--Reid correction term
does not easily apply, we set the entire correction term
to zero when (\ref{eq:condTau}) takes place for some $\psi_0$.
The correction term should be  especially important
when there are few studies and the heterogeneity 
between them is large.  

We can consider a variation of \eqref{eq:ffREpsi} in the case 
of the individual sources have few measurements, which 
means that the $\sigma_j$ estimates become uncertain. 
The II-CC-FF can then provide more sophisticated solutions. 
In the II step, we have exact CDs for each $\psi_j$ based on 
the Student's t distribution, which we can convert to 
a confidence log-likelihood by exact conversion in the CC step. 
For the FF step, we use the general random effect method 
from \eqref{eq:ffrandom}, either with numerical integration 
or using the TMB package. Corrections in both the II and 
FF step may be considered, but we have not fully investigated these 
options yet.

Rather than focusing on the overall mean, as we did above,
one may be interested in the overall spread $\tau$.
From the above, we have direct and Cox--Reid corrected
log-likelihood profiles
$\ell_{\fus,\prof}(\tau)=-\half A_k(\tau)$ and 
$\ell_{\fus,\cprof}(\tau)=-\half B_k(\tau)$, with 
\beqn
A_k(\tau)=\sumjk \Bigl[\log(\sigma_j^2+\tau^2)
   +{\{y_j-\hatt\psi_0(\tau)\}^2\over \sigma_j^2+\tau^2}\Bigr]  
\quadandquad
B_k(\tau)=A_k(\tau)+\log\Bigl(\sumjk{1\over \sigma_j^2+\tau^2}\Bigr), 
\eeqn 
with the profiled ML estimator 
$\hatt\psi_0(\tau)=\sumjk \hatt\psi_j/(\sigma_j^2+\tau^2)
   /\sumjk 1/(\sigma_j^2+\tau^2)$ for each given $\tau$. 
One might recognise the $-\half B_k(\tau)$ 
as the log-likelihood associated with 
the so-called restricted ML or REML estimator for $\tau$,
see for instance \citet{langan2019}. 
The link between the Cox--Reid correction (and more general 
corrections) and the REML procedure has been known for some 
time, see \citet{durban2000} and also \citet{CoxReid1987},
but is possibly under-appreciated in the meta-analysis
literature.
The above leads to two deviance functions,
using the direct and the corrected log-likelihood profiles, 
\beqn
D_{k,\ml}(\tau)=A_k(\tau)-A_k(\hatt\tau_\ml)
\quadandquad
D_{k,\cml}(\tau)=B_k(\tau)-B_k(\hatt\tau_\cml),
\eeqn 
with $\hatt\tau_\ml$ and $\hatt\tau_\cml$ the minimisers
of $A_k(\tau)$ and $B_k(\tau)$. 
The distribution of $y_j-\hatt\psi_0(\tau)$ does not 
depend on the underlying $\psi_0$, which means that 
$D_{k,\ml}(\tau)$ and $D_{k,\cml}(\tau)$ have distributions
only depending on the candidate value $\tau$. Thus 
we have well-defined and exact confidence curves for $\tau$,
\beq
\label{eq:ccandccfortau}
\cc_\ml(\tau)=\Pr_\tau\{D_{k,\ml}(\tau)\le D_{k,\ml,\obs}(\tau)\}
\quadandquad 
\cc_\cml(\tau)=\Pr_\tau\{D_{k,\cml}(\tau)\le D_{k,\cml,\obs}(\tau)\}. 
\eeq
We make use of these confidence curve methods 
in Application \ref{subsection:skulls}. 
The Wilks type chi-squared approximation
works well for moderate to large values of $k$,
but not for $\tau$ small, which often might be the 
parameter region of primary interest, as for the 
application pointed to. Hence we need to compute
the two confidence curves via simulations, 
with a high number of $D_{k,\ml}(\tau)$ and $D_{k,\cml}(\tau)$ 
generated for each candidate value $\tau$. 

In the normal-normal set-up above, the parameters are orthogonal, 
but we suggest that one could use the simple Cox--Reid correction 
even if the model for the observations in each source 
is non-normal. Suppose we are in a setting 
like in \eqref{eq:ffrandom}, and assume that the random 
effect distribution is normal, with the parameter 
of main interest being the overall mean $\psi_0$. 
Inside each source we can have any (regular) model. 
In a simple normal model, the Cox--Reid correction when 
profiling out the variance $\tau^2$ would be equal to 
$-\log\{\hatt\tau(\psi)\}$. We propose to routinely use 
the following corrected likelihood profile construction 
in this type of random effect setting,
\beq
\ell_{\fus,c\prof}(\psi_0) 
   = \sumjk \Bigl\{\log\Bigl[\int \exp\{\ell_{\con,j}(\psi_j)\}
   \frac{1}{\hatt \tau(\psi_0)}
   \varphi\left(\frac{\psi_j-\psi_0}{\hatt\tau(\psi_0)}\right)\,\dd\psi_j\Bigr] 
   \Bigr\}+ \log\{\hatt\tau(\psi_0)\},
\label{eq:ffrandomCorr}
\eeq 
where $\hatt\tau(\psi_0)$ is the ML estimate of $\tau$ 
for each fixed $\psi_0$ value and $\varphi$ is the standard 
normal density. The orthogonality of $\psi_0$ and $\tau$ 
in the full (integrated) distribution does not necessarily hold, 
but it would hold if the sources were large, since then 
$\ell_{\con,j}(\psi_j)$ above will approach a normal likelihood.  
We therefore consider formula \eqref{eq:ffrandomCorr} 
to be an approximate correction for the situation when $k$ 
is small, but the $n_j$ are sufficiently large. In fact the correction 
term in \eqref{eq:ffrandomCorr} is a special case of the correction 
term in  \eqref{eq:ffREpsi} when the $\sigma_j$ go to zero (which 
of course corresponds to  $n_j$ increasing towards infinity). 
We investigate this idea in the situation 
of meta-analysis of $2 \times 2$ tables, see Section 
\ref{subsection:meta2x2sim}.

\subsection{Meta-analyses of $2 \times 2$ tables}
\label{subsection:meta2x2}

In meta-analyses of $2 \times 2$ tables,
 each study seeks to compare the probability of observing a certain 
 binary event in the control group and in the treatment group.
The counting variables $Y_{0,j}$ and $Y_{1,j}$ indicate how many 
patients have experienced the event in each group in study $j$. 
These variables are usually modelled as pairs of binomials,
$Y_{0,j} \sim\binom(m_{0,j}, p_{0,j})$ 
and $Y_{1,j} \sim\binom(m_{1,j}, p_{1,j})$, 
with subscript `1' indicating treatment and `0' control, 
and the  sample sizes in each group are denoted by $m_{0,j}$ and 
$m_{1,j}$.
The most common measure for the treatment effect is the odds 
ratio, or equivalently, the log odds ratio $\psi_j$. For that effect 
measure it is convenient to express the event probabilities in the 
control and treatment groups as 
 $p_{0,j} = \exp(\theta_j)/(1+\exp(\theta_j))$,  
and $p_{1,j} = \exp(\theta_j +\psi_j)/(1+\exp(\theta_j+\psi_j))$.  
Each study, or source, has a specific nuisance parameter 
$\theta_j$, 
governing the event probability in the control group.
We will first treat 
the fixed effect case where the log odds ratios are 
assumed common across all sources, 
$\psi_1=\dots=\psi_k=\psi$, 
before we come to the random effect case in the next 
paragraph. The information available in each source depends 
on the binomial sample sizes $m_{0,j}$ and $m_{1,j}$, 
and on the event probabilities. If the number of studies 
increases while the size of each study stays constant, 
it is known that the ML estimator 
is inconsistent \citep{Breslow1981}, 
and we can expect that the standard II-CC-FF method will 
not work well. 
Also, the simple Cox--Reid correction to the profile 
in each source is not immediately available because 
$\psi_j$ and $\theta_j$ are not orthogonal. However, 
there exists an optimal CD for the common $\psi$ based on 
the theory from 
Section \ref{subsection:torenilsmethod},
\beq 
C_{\opt}(\psi,\data) = \Pr_\psi(B_k>b \midd z_1,\dots,z_k) 
   + \half \Pr_\psi(B_k = b \midd z_1,\dots,z_k).
\label{eq:2x2opt}
\eeq
Here, $z_j = y_{0,j}+y_{1,j}$ and $B_k=\sumjk Y_{1,j}$. 
The CD is obtained by simulating the distribution of $B_k$ 
given $Z_1, \dots, Z_k$. The second part of 
\eqref{eq:2x2opt} is a half-correction. 
Note also that we similarly have an optimal CD for $\psi_j$ 
within each source,
\beq
C_{\opt,j}(\psi_j,y_{0,j},y_{1,j}) 
   = P_\psi(Y_{1,j} > y_{1,j} \midd z_j) 
   + \half P_\psi(Y_{1,j} = y_{1,j} \midd z_j). 
\label{eq:cdi2}
\eeq 
This CD is simple to compute as $Y_{1,j} \midd Z_j$ 
has an eccentric hypergeometric distribution.  Note that this CD is 
very closely related to the method known as Fisher's exact test 
\citep{fisher1954}.
Starting from \eqref{eq:cdi2} for each source in the II step,  
we can obtain an approximation to the optimal solution 
in \eqref{eq:2x2opt} which is faster to compute and 
also lends itself to a natural random effect extension, as we will 
see.
In the CC step, we use exact conversion to obtain the 
confidence log-likelihoods $\ell_{\con,j}(\psi_j)=\log g_j(y_{1,j},\psi_j)$, 
where 
\beq
g_j(y_{1,j},\psi_j)
   ={{m_{0,j}\choose z_j-y_{1,j}}{m_{1,j}\choose y_{1,j}}\exp(\psi_j y_{1,j})
   \over \sum_{u=0}^{z_j} 
   {m_{0,j}\choose z_j-u}{m_{1,j}\choose u}\exp(\psi u) } 
   \quad {\rm for\ }y_{1,j}=0,1,\ldots,\min(z_j,m_{1,j}) 
\label{eq:exhyp}
\eeq 
is the density function of the eccentric hypergeometric 
distribution. We sum these confidence log-likelihoods
to get $\ell_\fus(\psi) = \sumjk \ell_{\con,j}(\psi_j)$, 
find the ML estimate $\hatt\psi$ 
and the deviance, and use the Wilks approximation:
\beq
\cc^{*}(\psi,\data) 
   = \Gamma_1(2\{\ell_\fus(\hatt\psi) - \ell_\fus(\psi)\}). 
\label{eq:cdi3}
\eeq
Even though there is some level of approximation in this solution, 
it tends to work well, see Section \ref{subsection:meta2x2sim}.

From this approximate fixed effect approach we find a natural
extension to random effects.   
Assuming that the log-odd ratios from the different 
sources come from a common normal distribution, 
we have the following fusion log-likelihood for the overall parameters, 
\beq
\ell_{\fus}(\psi,\tau) 
   =  \sumjk \log \Bigl\{\int g_j(y_{1,j}, \psi_j) 
   \frac{1}{ \tau}\varphi\left(\frac{\psi_j-\psi}{ \tau }
   \right)\,\dd\psi_j\Bigr\},
\label{eq:2x2random}
\eeq 
where $g_j(y_{1,j},\psi_j)$ is the density function of the eccentric 
hypergeometric distribution, pointed to above. 
If we are mainly interested in $\psi_0$, we  
profile out $\tau$ and use the Wilks approximation. 
If the heterogeneity is big or the number of sources is small we 
add the approximate 
Cox--Reid correction $\log\hatt \tau(\psi)$ 
from \eqref{eq:ffrandomCorr} to the profile log-likelihood. 
In applications, we computed the integral using the TMB package.
This approach seems promising with good coverage properties
in simulations (Section \ref{subsection:meta2x2sim}).
   
\section{Other combination methods based on CDs}
\label{section:kina} 


There is a steadily growing literature 
on combination of information with CDs. 
Here we will briefly discuss methods by 
\cite{SinghXieStrawderman05} and \citet{LLX15}. 
These CD approaches are sometimes collected under the same 
umbrella, 
called Fusion Learning \citep{ChengLiuXie2017}. 

We start by discussing the approach of 
\cite{SinghXieStrawderman05}, 
valid when all confidence components relate to a common
focus parameter. Suppose that independent information sources 
$y_1,\ldots,y_k$ give rise to CDs
for the same parameter, say 
$C_1(\psi,y_1),\ldots,C_k(\psi,y_k)$. 
A general way of combining these into a single
overall CD has been 
proposed and worked with by 
\citet{SinghXieStrawderman05}, later on applied
in various contexts by \citet{XieSingh2011}, \citet{XieSingh13}, 
\citet{LLX14}, and others. The 
starting point is that under the true state
of affairs, the $\Phi^{-1}(C_j(\psi,Y_j))$ 
are independent standard normals, from the 
basic properties of CDs; 
here $\Phi(\cdot)$ again denotes the c.d.f.~for 
the standard normal. Hence 
$\sumjk w_j\Phi^{-1}(C_j(\psi,Y_j))$ is also 
standard normal, when the weights $w_j$ are 
such that $\sumjk w_j^2=1$. This again implies that 
\beq
\label{eq:kina}
\bar C(\psi,y)=\Phi\Bigl(\sumjk w_j\Phi^{-1}(C_j(\psi,y_j))\Bigr) 
\eeq 
is a CD for $\psi$,
using the combined dataset $y=(y_1,\ldots,y_k)$. 
The idea generalises to other basic distributions 
than the normal, but then the required convolutions
become less tractable. 

For the prototype situation associated with 
(\ref{eq:simple}), the individual CDs
take the form $C_j(\psi,y_j)=\Phi((\psi-y_j)/\sigma_j)$,
and the general (\ref{eq:kina}) recipe yields 
$$\bar C(\psi,y)=\Phi\Bigl(\sumjk w_j(\psi-y_j)/\sigma_j\Bigr). $$  
Some considerations then lead to the best of these 
linear combinations, with weights $w_j$ proportional to $1/\sigma_j$ 
and $\sumjk w_j^2=1$. This indeed agrees with
the standard method (\ref{eq:kina}). 

Recipe (\ref{eq:kina}) requires nonrandom weights $w_j$,
and these could in various cases be fruitfully 
taken as proportional to $1/\sqrt{m_j}$, with $m_j$ the
sample size associated with data source $y_j$. 
In many other situations the balance is more delicate,
however, perhaps demanding nonrandom weights, of the
type $\hatt w_j$ estimating an underlying optimal but 
not observable $w_{j,0}$. Problems worked with 
in \citet{LLX14} are of this type. 
In such cases recipe (\ref{eq:kina}) is not entirely
appropriate and is rather to be seen as an approximation,
associated with confidence intervals with 
approximate levels of confidence. 

The approach described above yields approximative solutions 
for the basic normal-normal random effect model, 
partly helped by the fact that the unconditional 
density in that case has an explicit normal form, 
$y_j \sim \N(\psi_0, \sigma_j^2 + \tau^2)$. 
It is not clear how the method in \citet{XieSingh2011} 
can incorporate more general random effect models,
however. 

Under the `Fusion learning' umbrella there are other methods. 
The method in \citet{LLX15} may be termed a 
`confidence density method' and can be considered as 
a special case of II-CC-FF, as we will see. The method 
is proposed for a fixed effect setting, but where 
the studies may differ in reported outcomes, in measured 
covariates, or have source-specific nuisance parameters. 
Thus, some of the studies may only contain indirect 
information about the parameter of interest. Let $\theta$ 
be the full parameter vector for all the studies and 
$\gamma_j=M_j(\theta)$ the parameters in study $j$, 
with $M_j$ denoting a known mapping function.
\citet{LLX15} summarise the information in each source with 
multivariate normal CDs, $C_j(\gamma_j,y_j)$, 
transform these to confidence densities 
$c_j(\gamma_j,yj)=\partial C_j(\gamma_j,yj)/ \partial \gamma_j$,  
which are then multiplied into a combined confidence density, 
which informs on the full $\theta$. The authors stress 
that the approach is general in the sense that 
it can be used with a wide range of parametric models 
for the sources. This generality is achieved because 
the authors assume that the number of observations in each source 
increases to infinity. 
 
The normal CDs for each study only requires the estimated 
parameter vector $\hatt \gamma_j$ and estimated covariance 
matrix $\hatt \Sigma_j$ for $\hatt \gamma_j$, and the 
authors therefore highlight that the approach only 
needs summary statistics rather than the full data. 
Also, they prove that their approach is asymptotically 
equally efficient as a traditional likelihood approach 
using the full data.  

For location parameters in normal models the confidence density 
and confidence likelihood are proportional. The approach 
in \citet{LLX15} can therefore be considered a special case 
of  II-CC-FF. Sometimes the confidence density might be easier 
to obtain than the exact confidence log-likelihood, 
and could be used also in connection with II-CC-FF. However, 
one might need to be careful, as this approach could 
introduce mistakes. The confidence density is equal to 
$\dell C(\psi, T)/\dell\psi$, while the exact confidence 
likelihood takes the derivative with respect to $T$, 
as we saw in Section \ref{section:CC}. 
The difference between the confidence density 
and confidence likelihood will be the most pronounced 
when the sample sizes are small, with the difference
going away with increasing sample sizes. 

\section{Performance evaluations}
\label{section:performance}

The main benefit of our II-CC-FF framework is its general nature 
and wide applicability, and in the next section we will therefore 
demonstrate the use of II-CC-FF in several  non-standard 
combination situations.
Still, we also require that methods coming out of II-CC-FF should 
be competitive against other methods from the literature when 
applied to typical combination situations, for example 
meta-analysis settings. Here we will study 
the performance of II-CC-FF methods in two very
classical situations: 
first in the basic random effect model 
and then in the meta-analysis of $2\times 2$ tables. 
Both of these types of meta-analyses were discussed 
in Section \ref{section:meta}, along with some notation 
which will be used in this section too.
Finally, in the last subsection we will study 
confidence risk functions for some II-CC-FF schemes,
compared to other CD combination methods,
 in a particular simplified example. 

\subsection{The basic random effect model}
\label{subsection:basicREsim}

We have investigated some of the methods from Section 
\ref{subsection:basicRE} with a simulation study inspired by 
\citet{langan2019}. We treat the setting where $\psi_0$ is the 
parameter of main interest. In that setting \citet{langan2019} found 
that confidence intervals computed by the Hartung-Knapp-Sidik-
Jonkman method were the clear winners in terms of coverage 
properties. That method uses the traditional inverse-variance 
method for the point estimator, and computes confidence intervals 
based on the t-distribution and the following variance formula,
$$\Var_{\rm HKSJ}\,\hatt \psi_0
= \frac{\sumjk (\hatt \psi_j - \hatt \psi_0)^2/
   (\hatt\sigma_j^2+\hatt\tau^2)}{(k-1) 
   \sumjk 1/(\hatt\sigma_j^2+\hatt\tau^2)}. $$  
This formula requires one to plug-in an estimator 
for $\tau$ and for this we used the REML estimator, 
as recommended by \citet{langan2019}.
We will compare this state-of-the-art method with our standard 
II-CC-FF method, and the II-CC-FF method using the corrected 
log-likelihood profile in \eqref{eq:ffREpsi}. We will also include the 
method for the basic random effect model which comes out of the 
CD combination framework of \citet{XieSingh2011}. This method 
is implemented in the {\tt gmeta} package \citep{gmeta2016}. See 
Appendix \ref{section:appendixB} for more details.

We let $\psi_0=0.5$ and consider two values for the 
between-study heterogeneity, $\tau= 0.09$ for a scenario with little 
heterogeneity and  $\tau= 0.44$ for a scenario with large 
heterogeneity. Further, we choose $\sigma_j=2/\sqrt{m_j}$ with 
$m_j \sim \unif(30,50)$, and investigate four values for the 
number of studies, 5, 10, 20, 50. We generated 10000 
meta-analyses for each $k$ value. For a single meta-analysis,
$\psi_j$ and $\sigma_j$ were estimated using their ordinary 
estimators in each of the $k$ studies (even though all the 
methods assume that the $\sigma_j$ are known). For each method, we 
recorded the coverage rate of 95\% confidence intervals, the width 
of these intervals, and point estimates for $\psi_0$.
\begin{figure}[h]
\centering
\includegraphics[scale=0.45]{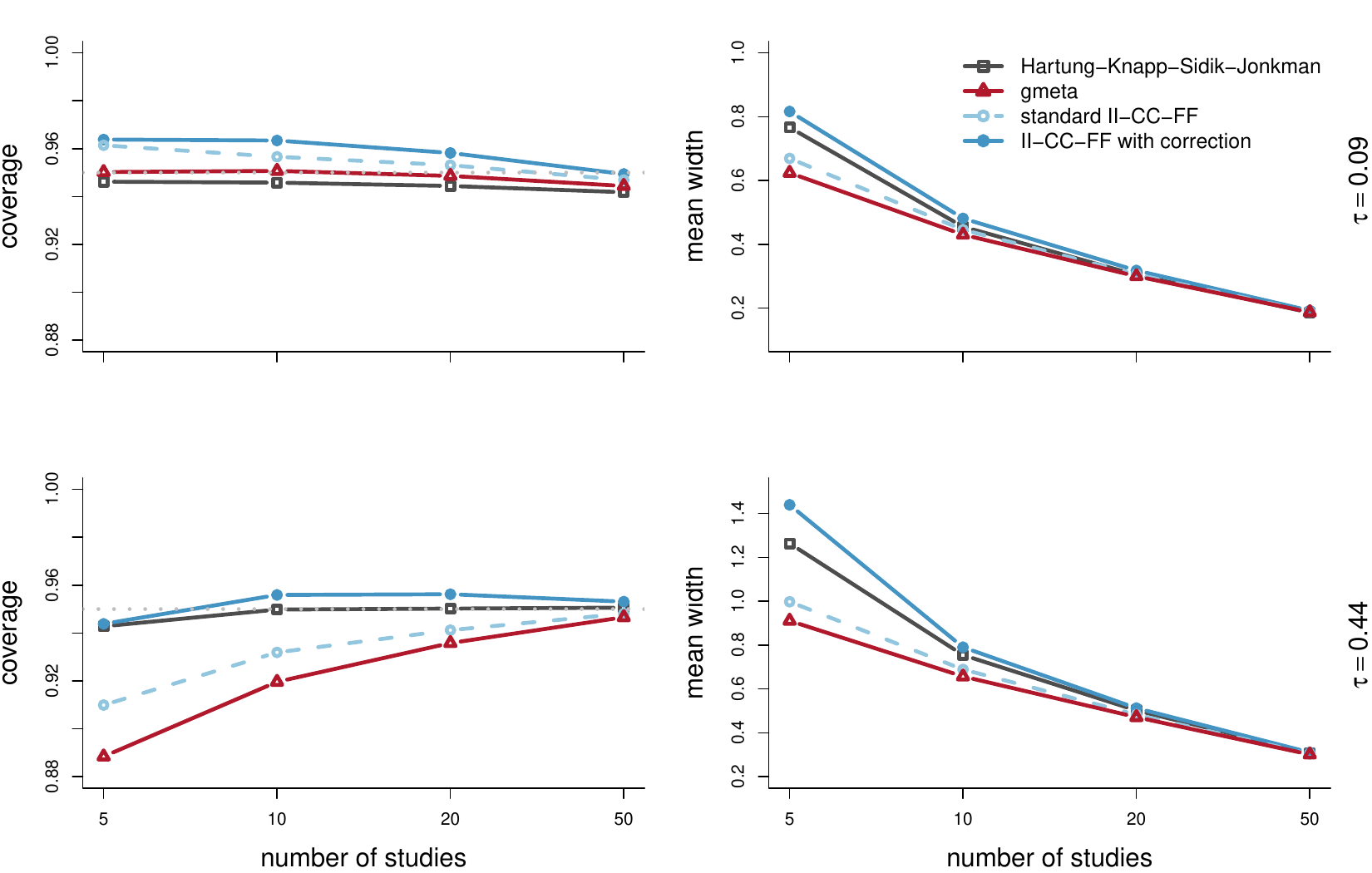}
\caption{Simulation results for the basic random effect model. The 
left plot gives the realised coverage rate 
of 95\% confidence intervals, the right plot gives the median width 
of these intervals. The results in the top row are for a scenario with 
small between-study heterogeneity, while in the bottom row the 
between-study heterogeneity is large. }
\label{figure:sim3}
\end{figure}

The confidence intervals from the Hartung-Knapp-Sidik-Jonkman 
method have  perfect coverage rate for almost all $k$ values in 
both scenarios (Figure \ref{figure:sim3}), which is consistent with 
results in \citet{langan2019}. The CD method from 
\citet{XieSingh2011} has very good performance in the scenario 
with small heterogeneity, but severe under-coverage in the 
scenario with large heterogeneity.  The standard II-CC-FF method 
has a similar problem in that scenario, but the II-CC-FF version 
with corrected profile likelihood obtains coverage rates close to the 
nominal 0.95.  Both II-CC-FF versions are slightly conservative in 
the small heterogeneity scenario, but obtain nonetheless 
confidence intervals of similar mean width as the 
Hartung-Knapp-Sidik-Jonkman intervals. The four methods 
produce virtually identical point estimates for $\psi_0$.
All in all we find the II-CC-FF method to have almost equally good 
performance as the state-of-the-art method. In scenarios with low 
heterogeneity the correction to the profile likelihood is not 
important, while it is crucial in scenarios with large between-study 
heterogeneity.

\subsection{Meta-analyses of $2 \times 2$ tables}
\label{subsection:meta2x2sim}

We will investigate both the fixed effect and random effect cases.
Our simulation set-ups are inspired by two recent papers with 
extensive simulation studies: \citet{piaget2019} for the fixed effect 
case 
and \citet{jackson2018} for the random effects case. 
With fixed effects, we will investigate three common effect 
measures: the (log) odds ratio defined in Section \ref{section:meta}, 
the (log) risk ratio, $\exp(\psi_j) = p_{1,j}/p_{0,j}$, 
and the risk difference, $\psi_j = p_{1,j}-p_{0,j}$. 
With random effects, we will only treat the odds ratio, however.

\citet{piaget2019} focus on meta-analyses with rare events, 
i.e.~where both the treatment and control group have low event 
probabilities and where there may be many studies with zero 
events.  
The overall conclusion of the paper was that the 
Mantel-Haenszel method had the best performance among the 
methods that were considered. This method is applicable to all 
three effect measures mentioned above, and \citet{piaget2019} 
found that it produced confidence intervals with good coverage 
properties, and point estimates with relatively small bias. The 
Mantel-Haenszel is a well-established  method, which was 
originally proposed in 1959 (\citet{mantel1959} and extended in 
\citet{rothman2008}). The method offers explicit estimators for the 
log odds ratio, log risk ratio and risk difference, as well as 
expressions for the variance of these estimators. Confidence 
intervals are computed using the Wald approximation. 
We will compare some variants of our II-CC-FF scheme with the 
Mantel-Haenszel method. 
We will also include methods reviewed in \citet{liu2014}. These 
methods are based on exact tests and  fall into the unifying CD 
framework of \citet{SinghXieStrawderman05} which we described in 
Section \ref{section:kina}. For the odds ratio and risk difference, 
these CD methods have been implemented in 
the {\tt gmeta} package \citep{gmeta2016}; for the risk ratio 
a related exact method is implemented in the {\tt exactmeta} 
package \citep{exactmeta2014}. 
See Appendix \ref{section:appendixB} for more details. 

For all three effect measures we will make use of the standard 
II-CC-FF procedure which consists of profiling out the 
nuisance parameters in each source, summing the log-likelihood 
profiles ,and then using the Wilks approximation to obtain a 
confidence curve for $\psi$, from which we can extract confidence 
intervals and point estimates. For the odds ratio case we will 
include two additional variants that can be said to fall under 
the II-CC-FF umbrella:  the optimal CD method given in 
\eqref{eq:2x2opt} and the II-CC-FF method using exact 
conversion, which 
we give in \eqref{eq:cdi3}. This II-CC-FF version resembles the 
standard II-CC-FF procedure, but avoids the profiling step by 
making use of the conditional log-likelihood for $\psi_j$, in 
\eqref{eq:exhyp}.

We simulated datasets with a median event probability of 0.005 in 
the control group. According to the simulation study in 
\citet{piaget2019}  we can expect that this will be a challenging 
setting for all methods we consider. 
We generated baseline probabilities 
$p_{0,j} = \exp(\theta_j)/\{1+\exp(\theta_j)\}$ by drawing 
$\theta_j \sim \N(\log \frac{0.005}{1-0.005},0.5)$, 
$m_{1,j} \sim \unif(50,150)$ and $m_{0,j} = r_j m_{1,j}$ with 
$r_j \sim \unif(0.5,1.5)$ 
(so we have some variation in which of the groups has the most 
participants). 
The probabilities in the treatment group where computed 
according to the chosen effect measure,
\beqn
& &\text{Odds ratio}\colon \qquad  
p_{1,j} = \frac{\exp(\theta_j+\psi)}{1+\exp(\theta_j+\psi)} 
   \quad \text{with the log odds ratio } \psi = -1.5,\\
& &\text{Risk ratio}\colon \qquad p_{1,j} = \exp(\psi) p_{0,j} 
   \quad \text{with the log risk ratio } \psi = -1.5,\\
& &\text{Risk difference}\colon \qquad p_{1,j} = \psi + p_{0,j} 
   \quad \text{with the risk difference } \psi = 0.05.
\eeqn 

We let $k$, the number of studies, take the values 5, 10, 20 and 
50. For each set-up we generated 10000 meta-analyses, except in 
the risk difference set-up where we only generated 1000 (because 
the {\tt gmeta} package was extremely slow for this effect 
measure). For each method we computed the coverage rate of 95\% 
confidence intervals, the median width of these intervals, and the 
median bias of the point estimates coming out of the methods. 
Both the odds ratio and risk ratio where analysed on the log scale.

\begin{figure}[h]
\centering
\includegraphics[scale=0.55]{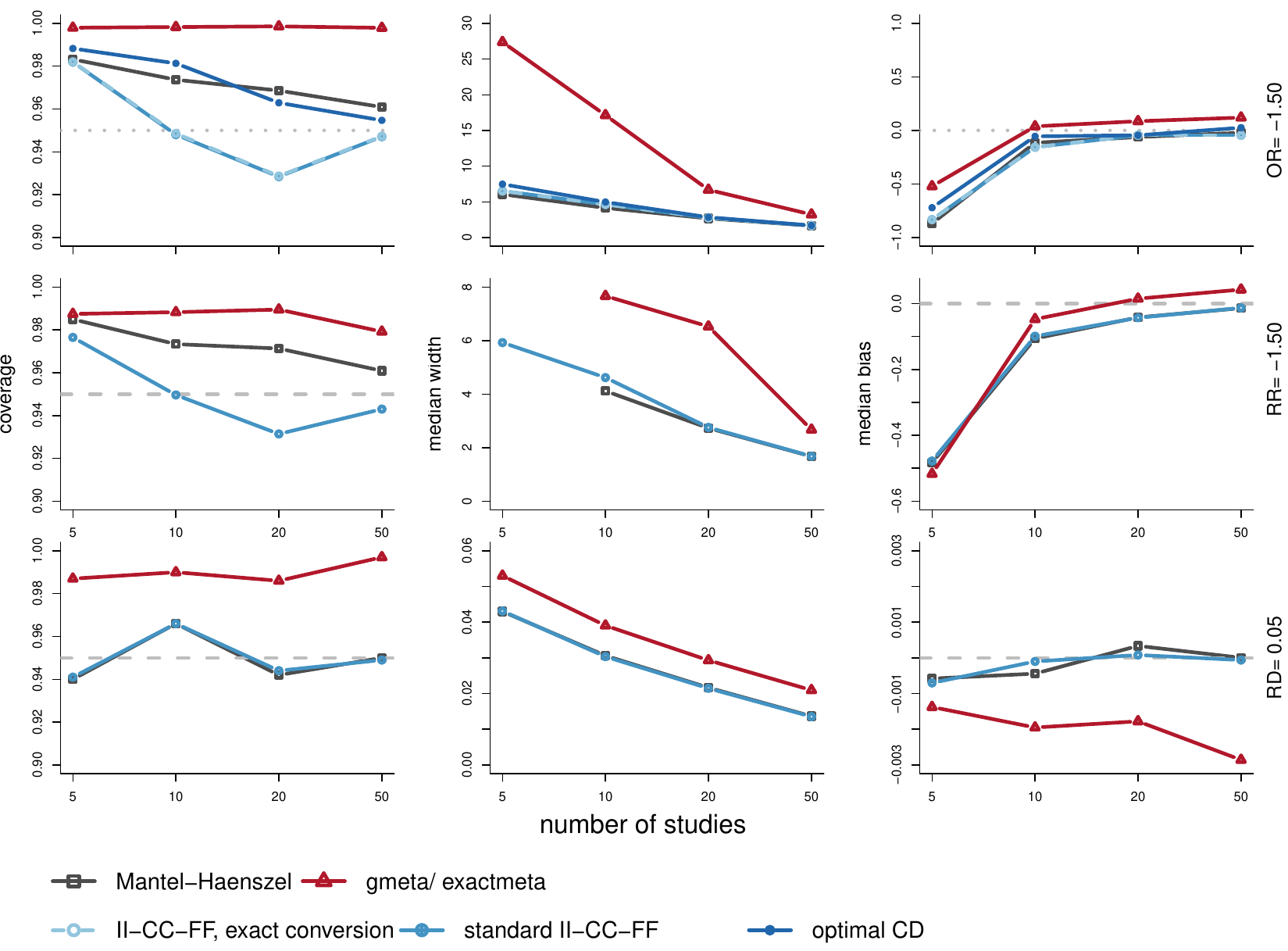}
\caption{Simulation results for fixed effect meta-analysis of 
$2 \times 2$ tables. The left column gives the realised coverage 
rate of 95\% confidence intervals, the middle column gives the 
median width of these intervals and the right column gives the 
median bias of the point estimate coming from each of the 
methods. The top row gives the results for the (log) odds ratio, the 
middle row for the (log) risk ratio and the bottom row gives the 
results for the risk difference. When $k$ is small the confidence 
intervals can sometimes have infinite width (which explains the 
missing points for the log risk ratio).}
\label{figure:sim1}
\end{figure}

Results are presented in Figure \ref{figure:sim1}.  
When there are few studies, most method struggle with 
over-coverage in the odds ratio and risk ratio set-ups, but the 
Mantel-Haenszel method and the various II-CC-FF schemes come 
close to the nominal level as $k$ increases. 
 For the odds ratio, the standard II-CC-FF and the II-CC-FF with 
 exact conversion have practically identical performance: both have 
 some degree of under-coverage for $k=20$, but come very close 
 to the nominal 0.95 for $k=50$. This seems to indicate that even 
 though we are in a rare events setting, there is nonetheless 
 sufficient information within each source so as to safely profile out 
 the $\theta_j$. The under-coverage experienced by these methods 
 for $k=20$ indicate that the Wilks approximation is not perfectly 
 fine, and one could consider adjustments, 
like for instance the Bartlett correction. 
The standard II-CC-FF has reasonably good coverage properties for 
the risk ratio and risk difference. For the risk ratio there is again 
some degree of under-coverage for $k=20$, but for the risk 
difference there is no such pattern. In that setting, the Mantel-
Haenszel method and the standard II-CC-FF method have 
practically identical performance. The {\tt gmeta} and  
{\tt exactmeta} methods are very conservative in all these 
experiments and consistently produce very wide 95\% intervals, 
with a coverage rate close to 1. 
When there are few events, all methods tend to underestimate the 
effect measure (giving a negative median bias). As $k$ increases all 
methods come closer to the true effect, except for the {\tt gmeta} 
method for the risk difference which appears to get worse. For the 
odds ratio and risk ratio however, the {\tt gmeta} method produces 
point estimates with somewhat smaller median bias than the other 
methods for some values of $k$. 

Overall, II-CC-FF performs similarly, and at best, slightly better 
than the main existing competitor (according to \citet{piaget2019}). 
The II-CC-FF and the Mantel-Haenszel methods have generally 
similarly wide confidence intervals, but II-CC-FF intervals have a  
coverage rate often coming closer to the nominal level. Sometimes 
the standard II-CC-FF method produces intervals with some degree 
of under-coverage, but the optimal CD method (which is only 
available for the odds ratio) does not.
The methods reviewed in \citet{liu2014} perform very poorly in the 
settings we have presented here. We have included a simulation 
study with less extreme event probabilities in the appendix. There 
the median event probability in the control group is 0.1. That 
situation is less challenging for all the methods, and most obtain a 
coverage rate close to 0.95 even for small $k$. The {\tt gmeta} 
method has acceptable performance in that setting, at least for the 
odds ratio. For the risk difference the results are still quite poor.

Some readers might be puzzled by the fact that the optimal CD 
method for the odds ratio does not have even better performance 
in the simulations. This CD is optimal in the sense of 
Section~\ref{subsection:torenilsmethod} and one might think that it 
should have exact coverage properties. For discrete data 
as we have here, the method of Section~\ref{subsection:torenilsmethod} 
requires a half-correction, as we see in \eqref{eq:2x2opt}. 
In settings with few tables and few events, 
the statistics encountered will attain only few possible values,
and so have particularly non-continuous distributions; 
athis likely explains the results we see in these simulations. 

Finally, note that in rare events settings it might be fruitful to 
assume that the event counts in the two groups are  Poisson 
distributed rather than binomials, see for instance \citet{cai2010} 
and \citet{CunenHjort15}. Our II-CC-FF framework can naturally be 
applied to such models too and would produce different results 
than the ones we see here. 

In \citet{jackson2018} the authors compare seven methods for 
odds ratio meta-analysis of $2 \times 2$ tables in a random effects 
setting. 
Many authors find random effects methods, and implicitly random 
effects models, to be preferable to fixed effects methods for this 
type of meta-analysis, since it may seem more realistic to allow for 
some heterogeneity in the treatment effects between the studies. 
As is commonly done, we will assume that the log odds ratios come 
from a normal distribution, $\psi_j \sim \N(\psi_0,\tau^2)$.
In their simulation study, \citet{jackson2018} found that several of 
the methods had similar (equally good) performance. Among 
others, they recommend the use of a modified version of the  
\citet{simmonds2016} method, for its ease of use and good 
performance. 
The method uses a generalised linear mixed effect framework to 
estimate $\psi_0$ and $\tau$ (and the $\theta_j$) by assuming 
that the logit of $(p_{0,j},p_{1,j})$ are drawn from a bivariate 
normal distribution with expectation $(\theta_j, \theta_j + \psi_0)$ 
and a certain covariance matrix (with variances equal to 
$\tau^2/4$).
The modified Simmonds-Higgins method is implemented in the 
{\tt metafor} package \citep{metafor2010}, and we will  compare 
our II-CC-FF methods with this.  The other CD methods we 
reviewed in Section \ref{section:kina} do not provide a readily 
applicable method for these type of random effects situations.

We will investigate two II-CC-FF versions here. The first version 
constitutes the standard II-CC-FF solution for random effects. 
We use \eqref{eq:2x2random},  profile out $\tau$ and use the 
Wilks approximation to obtain the combined confidence curve for 
$\psi_0$. The second version makes use of the same profile 
log-likelihood, but adds the approximate Cox--Reid correction as 
suggested in the text under \eqref{eq:2x2random}. The idea is that 
this correction will account for the error from having profiled out 
$\tau$. We  compute the integral in \eqref{eq:2x2random}  using 
the TMB package.  Note that we do not compute the correction in 
rounds where $\tau$ is estimated to be very close to zero (below 
0.0001), since in that case the correction blows up. 

\citet{jackson2018} include many different scenarios in their 
simulations, but we limit ourselves to one main scenario, where we 
let the number of studies, $k$, vary as we did in the previous 
subsection. We use  
$\theta_j \sim \N(\log \frac{0.2}{1-0.2},0.3^2)$ giving median 
baseline probabilities of 0.2, $m_{1,j} \sim \unif(10,50)$ and 
$m_{0,j} = m_{1,j}$. Further, we let $\psi_j \sim \N(0,0.168)$, 
which gives considerable heterogeneity in the treatment effects. We 
generated 10000 meta-analyses for each value of $k$ (5, 10, 20 
and 50).

\begin{figure}[h]
\centering
\includegraphics[scale=0.45]{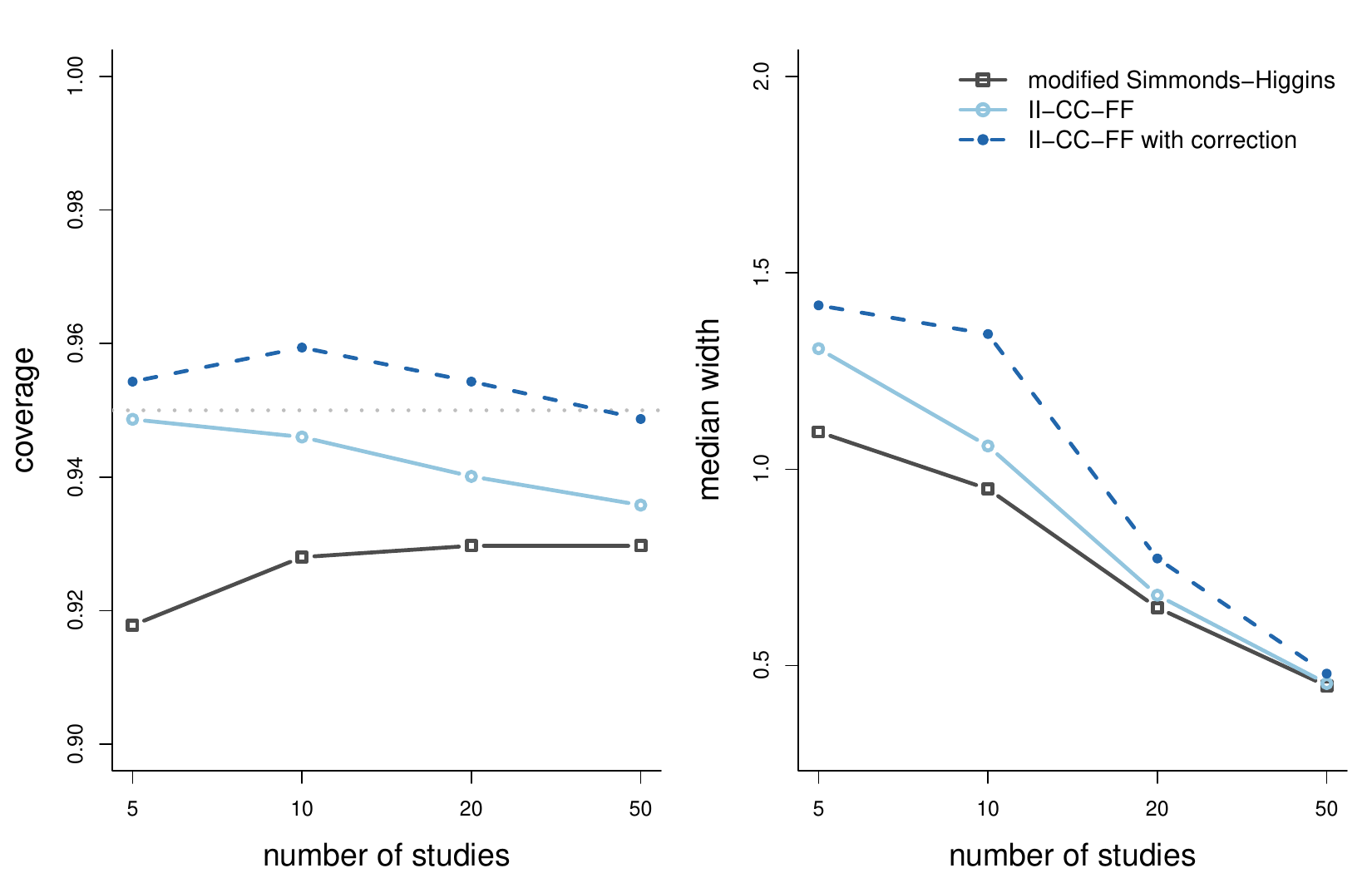}
\caption{Simulation results for random effect meta-analysis of 
$2 \times 2$ tables. The left plot gives the realised coverage rate 
of 95\% confidence intervals, the right plot gives the median width 
of these intervals. }
\label{figure:sim2}
\end{figure}

For each method we computed the coverage rate of 95\% 
confidence intervals, the median width of these intervals, and we 
recorded the point estimates coming out of the methods. In 
Figure \ref{figure:sim2} we display the coverage rate and median 
width results.
Both the modified Simmonds and Higgins and standard II-CC-FF 
methods produce confidence intervals with some degree of under-
coverage. For  the modified Simmonds and Higgins method this is 
consistent with the results in \citet{jackson2018} for scenarios with 
considerable heterogeneity in the treatment effects and small 
within-study sample sizes (as we have here).  Surprisingly, the 
coverage rate of the intervals from the standard II-CC-FF seems to 
worsen as $k$ increases. The performance of the corrected
II-CC-FF method in terms of coverage rate is very good. The effect 
of the correction seems substantial even when $k$ is quite large. 
This is somewhat unexpected, but probably due to the relatively 
large $\tau$ value in this scenario. For $k=50$, the median widths 
of confidence intervals from the three methods are almost 
identical. If we had displayed the mean widths instead, we would 
have seen that the corrected II-CC-FF method in fact has larger 
\textit{mean} width, which explains the higher coverage rate.
Note that the Cox--Reid correction term in this case usually only 
widens the confidence curve, and only rarely shifts the point 
estimates. 

The three methods produce very similar estimates of $\psi_0$ and 
$\tau$. The bias for $\psi_0$ is small, but all three methods tend 
to under-estimate $\tau$, which is a feature they share with all the 
methods investigated in \citet{jackson2018}. Remember that the 
II-CC-FF methods 
are in this case focussing on $\psi_0$, and if $\tau$ had 
been of primary interest we would have used the confidence curve given in \eqref{eq:ccandccfortau}.

In Appendix \ref{section:appendixB} we include a similar figure for a scenario with less 
heterogeneity and higher within-study sample sizes. In that case, 
the modified Simmonds-Higgins method has close to correct 
coverage rate for all values of $k$, while both II-CC-FF methods 
are slightly conservative. When $\tau$ is small the correction term 
seldom changes the confidence curve, and the two II-CC-FF 
confidence curves are therefore often very similar in that scenario. 

Again, we find II-CC-FF to have an overall good performance. In 
situations with large heterogeneity the corrected II-CC-FF method 
outperforms the recommended method from the literature.
The scenarios we have investigated here have quite large event 
probabilities, and it is conceivable that the results would be 
somewhat different in a rare events setting. We hope to investigate 
this issue further in a separate paper. 

Our standard II-CC-FF 
solution has actually been suggested previously, 
in \citet{Stijnen2010}, and also in \citet{van1993}, as the 
hypergeometric-normal model. This model is in fact among the 
seven studied in \citet{jackson2018}, and while it obtains 
reasonably good performance in their simulations, the authors 
report numerical problems and estimation failure. This could be 
related to a different implementation than ours and particularly to 
the Laplace approximations used in the TMB package. We did not 
find that the standard II-CC-FF had a high probability of failure in 
the scenarios we investigated, but we readily acknowledge that 
further  implementation efforts are necessary in order to make the 
method widely applicable. 
The approximate Cox--Reid correction which we found fruitful in 
settings where the standard II-CC-FF tended to produce 
anti-conservative intervals was not investigated in \citet{jackson2018}, 
or anywhere else in the literature as far as we know.

\subsection{Confidence risk in a prototype example}
\label{subsection:nils}

To see how versions of the II-CC-FF scheme
fare against natural competitors, in terms
of leading to accurate CDs for focus parameters,
consider the following simple setup. 
Independent gamma distributed variables 
$Y_j\sim\Gam(a_j,\theta)$
are observed for $j=1,\ldots,k$,
with densities proportional to $y_j^{a_j-1}\exp(-\theta y_j)$,
with known shape parameters $a_j$ and unknown
scale parameter $\theta$. Insights gleaned from 
studying performance in such a prototype setup should 
be helpful when analysing more complex situations.
Here the canonical CD for data source $y_j$ alone is
\beqn
C_j(\theta,y_j)
=\Pr_{\theta}\{Y_j\le y_j\}
=G(\theta y_j,a_j,1),  
\eeqn
with $G(\cdot,a_j,1)$ the c.d.f.~of the $\Gam(a_j,1)$ distribution. 
The II-CC-FF scheme, using the directly available
log-likelihood components, leads here to 
\beqn
\ell_\fus(\theta)
=\sumjk(a_j\log\theta-\theta y_j)
=a_{\cdot}\log\theta-\theta y_{\cdot}, 
\eeqn 
writing $a_{\cdot}=\sumjk a_j$ and $y_{\cdot}=\sumjk y_j$. 
From $y_{\cdot}\sim\Gam(a_{\cdot},\theta)$, its natural associated CD is
$C^*(\theta,y)=G(\theta y_{\cdot},a_{\cdot})$.
By the optimality theorem of Section \ref{subsection:torenilsmethod}, 
this CD uniformly outperforms all competitors,
in terms of all risk functions of the (\ref{eq:risk}) type.

Competitors to consider, for performance comparisons, 
include the following. 
(i) 
The CD method of \citet{SinghXieStrawderman05},
with the normal transformation, as in (\ref{eq:kina}),
with $C_j(\theta_j,y_j)$ as above, and with the natural choice
$w_j=(a_j/a_{\cdot})^{1/2}$.
(ii) 
The ML estimator is $\hatt\theta=a_{\cdot}/y_{\cdot}$,
and CDs may be constructed based on its exact or 
approximate disribution. The deviance function 
associated with the log-likelihood function is found to be 
\beqn
D_\fus(\theta)
=2\{\ell(\hatt\theta)-\ell(\theta)\}
=2\{a_{\cdot}\log(\hatt\theta/\theta)-(\hatt\theta-\theta)y_{\cdot}\}
=2a_{\cdot} \{\log(\hatt\theta/\theta)-(\hatt\theta-\theta)/\hatt\theta\}.
\eeqn 
Its distribution is independent of $\theta$.
Indeed, from $y_\cdot\sim\Gam(a_\cdot,\theta)$ one may
write $\hatt\theta\sim\theta/V$, in terms of
$V\sim\Gam(a_{\cdot},1)/a_\cdot$, and from this follows 
the representation $D=2a_{\cdot}(V-1-\log V)$. 
An exact confidence curve is therefore
\beqn
\cc_\fus(\theta,y)=H(D_\fus(\theta),a_\cdot), 
\eeqn 
writing $H(x,a_\cdot)$ for the c.d.f.~of the $D$.
For moderate to large $a_\cdot$, some analysis shows
$D\doteq a_\cdot(V-1)^2$, and in particular 
$H(x,a_\cdot)$ is then close to the chi-squared distribution
$\Gamma_1(x)$, in line with the Wilks theorem. 
Inspection shows that the $\chi^2_1$ approximation
works well already from say $a_\cdot\ge 6.0$, 
so only for smaller values is it worthwhile 
computing the exact $H(D_\fus(\theta,a_\cdot)$. 
The $a_\cdot$ factor needs to be bigger in order 
for the normal approximation based 
$\Phi(a_\cdot^{1/2}(1-\theta/\hatt\theta)$  
to come close to the $\Gamma_1(D_\fus(\theta))$, however. 
(iii) 
We also point to the `confidence density method'  argued for in \citet{LLX15},
in essence consisting in deriving the confidence densities
from the individual CDs, here taking the form 
$c_j(\theta)=g(\theta y_j,a_j,1)y_j$,  
and then treating the resulting
$c_1(\theta)\cdots c_k(\theta)$ as a likelihood function. 
In the present case, this is seen to be proportional to 
$\theta^{\sumjk(a_j-1)}\exp(-\theta y_\cdot)$, 
leading to the estimator $\tilda\theta=(a_\cdot-k)/y_\cdot$,
which has a sometimes severe negative bias. 
This serves to note that approximate CDs 
coming from the general recipes associated 
with combination of confidence densities
may not work well, without further fine-tuning. 
It does work well in the multinormal setups
studied in \citet{LLX15}, where confidence densities 
become identical to the converted log-likelihoods,
but not in general. 

We may illustrate the general performance theory 
by using $\Gamma(u)=|u|$ in (\ref{eq:risk}), 
so risk is measured by the smallness of 
$\E_\theta\,|\theta_\cd-\theta|$. Again 
$\theta_\cd$ is a random draw from the CD in question, 
which is itself the result of random data. 
For the best method $C^*(\theta,y)$, 
some analysis shows that $r^*(\theta)=r_0\theta$,
with $r_0=\E\,|G_1/G_2-1|$, where $G_1$ and $G_2$
are independent draws from the $\Gam(a_{\cdot},1)$
distribution. 
The methods based on the $\ell_\fus(\theta)$,
as both this optimal $C^*(\theta,y)$ and 
those using the deviance, offer sometimes drastic 
improvements over both the \citet{LLX14, LLX15} methods. 
for either small or big $\theta$, depending on both $k$
and the sizes of the $a_j$. 
The improvement is most noticeable in cases with
many groups and small $a_j$.
There is no simple expression for the risk $r(\theta)$
for the \citet{SinghXieStrawderman05} method, but 
it may be computed numerically by simulating for each 
value of $\theta$ a high number of $|\theta_\cd-\theta|$ 
in the natural two-stage fashion; first data $y$ 
from the model, leading to the CD $C(\theta,y)$, 
then $\theta_\cd$ from this distribution.

\section{Applications}
\label{section:applications}

Below we illustrate the capacity for the II-CC-FF paradigm
to solve problems in four rather different application settings. 
The first application concerns an interesting 
archaeological dataset. Here we use the so-called basic 
random effect model which was discussed in 
Sections \ref{subsection:basicRE} and \ref{subsection:basicREsim}, 
but perhaps atypically our parameter of main interest 
is the spread parameter, not the overall centre parameter. 
For this spread parameter we construct exact CD methods 
for the FF step.
The annual growth rate of humpback whales is 
the focus of our second application story. There
we illustrate how to construct confidence curves based
on non-sufficient summary statistics; we only have access
to a point estimate and a highly non-symmetric confidence interval. 
In this example we also demonstrate how partial prior 
information can be incorporated into our II-CC-FF framework. 
Our third story concerns the development over time 
of the median Body Mass Index for Olympic speekskaters, 
where part of the the challenge is to construct and 
then convert accurate nonparametric CDs for sample medians 
to parametric log-likelihood terms. 
Finally, the last application illustrates the combination 
of `hard' with `soft' data. Here `hard' designates 
data sources of high quality which inform directly 
on the focus parameter. `Soft' data, on the other hand, 
may be of lower quality, with more noise and biases, 
or simply containing less direct information on the focus parameter. 
Such large, noisy datasets are increasingly available 
in a number of fields, for example from webscraping 
or text-mining, but lead to challenges when attempting 
to fuse the sources. We illustrate the combination of `hard' 
and`soft' data with a grand question from the field 
of peace and conflict research; is there evidence for 
The long peace, and in that case, when did it start?

\subsection{Skullometrics} 
\label{subsection:skulls}

In their fascinating anthropometrical study of
the inhabitants of Upper Egypt, from the earliest prehistoric
times to the Mohammedan Conquest, \citet{ThomsonRandall05}
report on skull measurements for more than a thousand 
crania. A subset of their data is reported on and
analysed in \citet[Chs.~1 and 9]{ClaeskensHjort08},
see in particular their Figures 1.1 and 9.1. 
This pertains to four cranium measurements, say 
$y=(y_1,y_2,y_3,y_4)^\tr$, for 30 skulls, from each
of five Egyptian time epochs, corresponding to 
$-4000,-3300,-1850,-200,150$ on our A.D.~scale. 
We model these vectors as 
\beqn
Y_{j,i}\sim\N_4(\xi_j,\Sigma_j) 
   \quad \hbox{\rm for\ }i=1,\ldots,30, 
\eeqn
for each of the five epochs $j$. 
There is a variety of parameters worth recording
and analysing, where the emphasis is on identifying
the necessarily small changes over time,
related to the history of emigration and immigration
in ancient Egypt; see also \citet[Example 3.10]{CLP}. 
For the present illustration we choose to focus on 
the variance matrices, not the means, and consider 
\beqn 
\psi=\{\max\eigen(\Sigma)\}^{1/2}/\{\min\eigen(\Sigma)\}^{1/2},
\eeqn
the ratio of the largest root-eigenvalue 
to the smallest root-eigenvalue of the variance matrix
of the four skull measurements. This is the ratio 
of the largest to the smallest standard deviations 
of linear combinations $a^\tr Y$ 
of the four skull measurements, normalised to have 
coefficient vector length $\|a\|=1$. 
This parameter is one of several natural measures
of the degree to which the skull distribution is `stretched'. 
The question is whether the stretch parameter $\psi$ 
has changed over time. We assess the degree of change, 
if any, via the spread parameter $\tau$ in the natural 
model taking $\psi_1,\ldots,\psi_5\sim\N(\psi_0,\tau^2)$.
Rather than merely providing a test of the implied
hypothesis $H_0\colon\psi_1=\cdots=\psi_5$, which is equivalent
to $\tau=0$, with its inevitable p-value and a yes-no
answer as with a traditional one-way layout type test, 
we aim at giving a full CD
for $\tau$, again applying the II-CC-FF scheme. 

\begin{table}[h]
\caption{Skulls: For each of the five time epochs, 
the table gives the estimate $\hatt\psi$ and its 
estimated standard deviation $\hatt\sigma$. 
See Section \ref{subsection:skulls} and Figure \ref{figure:skulls}. 
\label{table:skulls}} 
\small
\begin{center}
\begin{tabular}{ccc}  
epoch &  $\hatt\psi$ &$\hatt\sigma$ \\ \midrule
$-4000$ &  2.652 &  0.561 \\ 
$-3300$ &  2.117 &  0.444 \\ 
$-1850$ &  1.564 &  0.331 \\ 
$ -200$ &  2.914 &  0.620 \\ 
$\phantom{-} 150$ &  1.764 &  0.373 \\ 
\end{tabular}
\end{center}
\end{table}

\begin{figure}[h] 
\begin{center}
\includegraphics[scale=0.4]{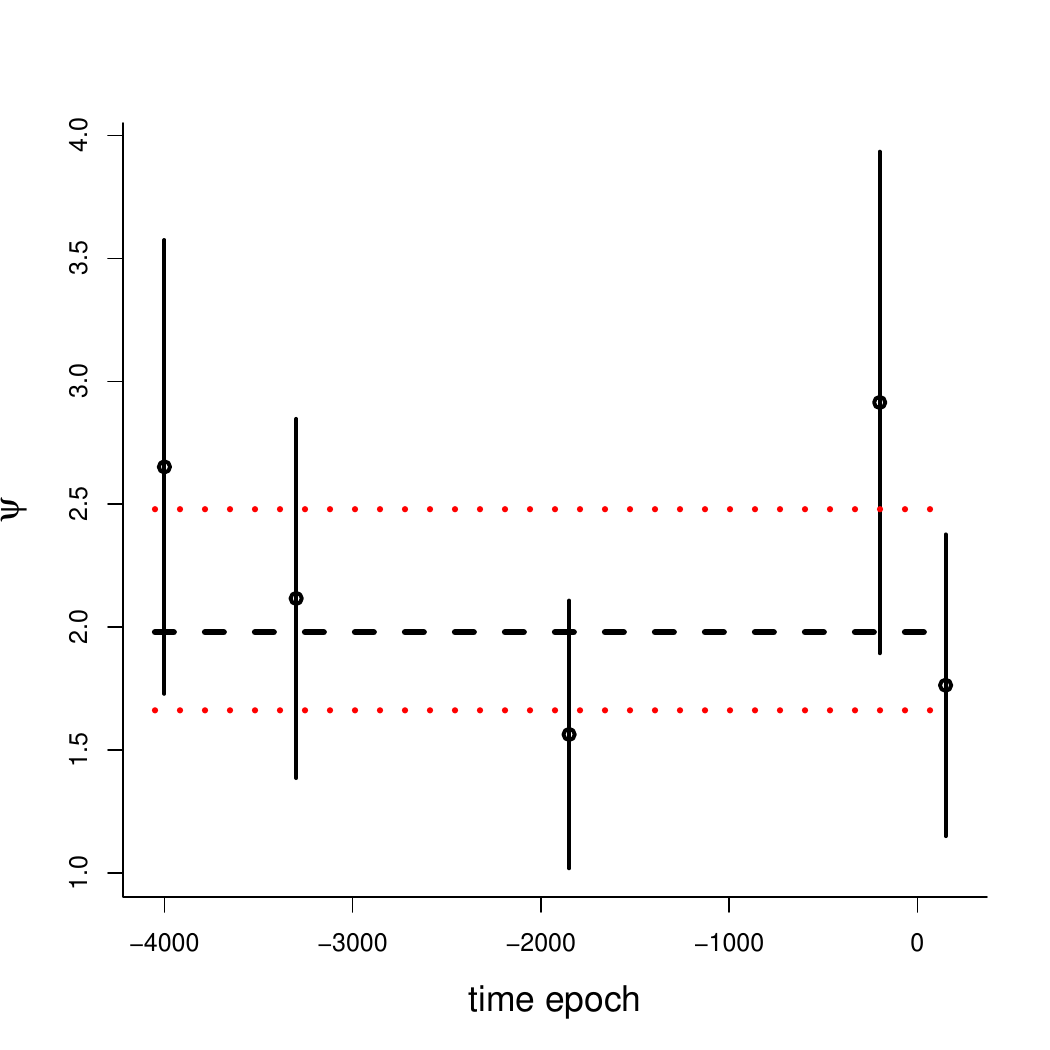}
\includegraphics[scale=0.4]{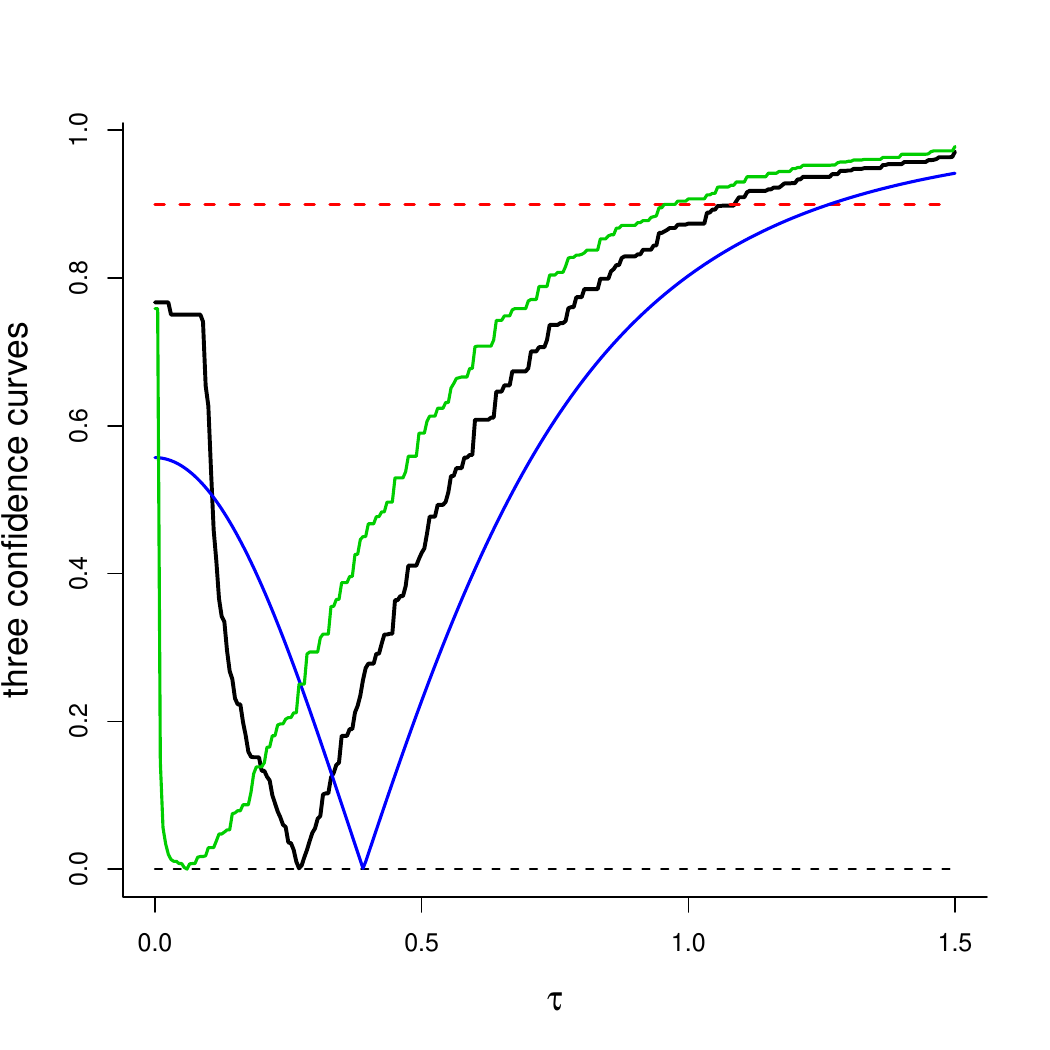}
\end{center}
\caption{Left panel: 
Point estimates $\hatt\psi_j$ with 90\%
confidence intervals, for the skull stretch
parameter $\psi$, across five time epochs
(see Table \ref{table:skulls}).  
Right panel: Three confidence curves for 
the spread parameter $\tau$, with median confidence
estimates 0.006 (the direct profile method),
0.272 (the corrected profile method), 
0.390 (the $Q_k(\tau)$ method).
\label{figure:skulls} }
\end{figure}

Table \ref{table:skulls} gives point estimates 
\beqn 
\hatt\psi_j=\{\max\eigen(\hatt\Sigma_j)\}^{1/2}
   /\{\min\eigen(\hatt\Sigma_j)\}^{1/2} 
\eeqn 
for the five time epochs, along with estimated standard
deviations $\sigma_j$ for these estimators, the latter obtained 
via parametric bootstrapping from the estimated multinormal 
distributions. For our present purposes the underlying 
distributions for the estimators are approximately normal, 
with the standard deviations $\sigma_j$ approximately known. 
Figure \ref{figure:skulls} displays point estimates
with 0.90 confidence intervals (left panel), for the five epochs. 

Using log-likelihood fusion methods derived
in Section \ref{subsection:basicRE}, 
involving profiling and corrections, we may
compute the confidence curves $\cc_\ml(\tau)$ and $\cc_\cml(\tau)$,
as per (\ref{eq:ccandccfortau}). These involve simulation
of a high number of deviance statistics
for each candidate value of $\tau$. 
The resulting confidence curves are shown in 
Figure \ref{figure:skulls} (right panel). The direct profile
method can be shown to have a clear negative bias,
particularly so for smaller values of $k$. 
For the present case of $k=5$ the corrected version 
$\cc_\cml(\tau)$, with median confidence estimate 0.272,
is better than the direct version $\cc_\ml(\tau)$,
with median confidence estimate 0.006. 
A third and simpler to compute CD for $\tau$ is via 
$$Q_k(\tau)=\sumjk {\{\hatt\psi_j-\hatt\psi_0(\tau)\}^2
   \over \sigma_j^2+\tau^2}
  \quad \hbox{\rm and} \quad 
  C(\tau, \data)=1-\Gamma_{k-1}(Q_{k,\obs}(\tau)), $$
the point being that $Q_k(\tau)$ for a given true 
value of $\tau$ has the $\chi^2_{k-1}$ distribution; 
see \citet[Ch.~13]{CLP}.
We note that these confidence curves have point masses at zero,
hence also the associated CDs, via $\cc(\tau)=|1-2C(\tau)|$. 
For each, the $C(0)$ has a clear interpretation as
the p-value for testing $\tau=0$ against $\tau>0$.
For the corrected profile CD, we find $C(0)=0.123$, 
not small enough to warrant a claim that this particular
$\psi$ parameter has changed over the four thousand years
of Egyptian history -- other skullometric parameters
have however changed; see \citet[Section 9.1]{ClaeskensHjort08} 
and \citet[Example 3.5]{CLP}. An accurate 0.90 interval 
for $\tau$, using the corrected profile, 
also indicated in the figure, is $[0,1.085]$,
with median confidence estimate is $0.272$. 

In other applications of this type of extended 
meta-analysis machinery the centre value $\psi_0$
of the background distribution of the $\psi_j$ might
be of high importance. 
For the skulls analysis the primary question is whether
the $\psi_j$ parameter, or other similar parameters 
associated with the $\Sigma_j$ matrices, have changed
over the course of four thousand years, and the precise 
value of $\psi_0$ is of secondary importance. 
We report, though, that the corrected log-profile methods 
of Section \ref{subsection:basicRE}, see (\ref{eq:ffREpsi}), 
lead to overall point estimate 1.980,
with an accurate 90\% interval stretching from 1.662 to 2.480.
These intervals are not symmetric around the point estimate; 
see left panel of Figure \ref{figure:skulls}.

\subsection{Abundance of humpback whales}
\label{subsection:whales}

The II-CC-FF paradigm readily lends itself to combination 
of information from published sources,
where we may not have access to the full data, but only 
summary measures. \citet{Paxtonetal09}
provide estimates of the abundance of humpback whales 
in the North Atlantic in the years 1995 and 2001. 
The two estimates are based on different surveys and 
can be considered independent. The authors also provide 95\% 
confidence intervals, via a somewhat complicated model
involving aggregation of line transect data from 
different areas via spatial smoothing, and also includes
bootstrapping. The available information is as
presented in Table \ref{table:whales}; note here that 
the natural 95\% confidence interval is not at all
symmetric around the point estimate, with 
an implied skewness to the right. 
 
\begin{table}[h]
\caption{Abundance assessment of a humpback population,
from 1995 and 2001, summarised as 2.5\%, 50\%, 97.5\% 
confidence quantiles; from \citet{Paxtonetal09}. 
See Section \ref{subsection:whales} and Figure \ref{figure:whales}.
\label{table:whales}}
\small
\begin{center}
\begin{tabular}{cccc}  
         & 2.5\% & 50\% & 97.5\% \\  \midrule
    1995 &  3439 & 9810 & 21457 \\ 
    2001 & 6651 & 11319  & 21214\\ 
\end{tabular}
\end{center}
\end{table}

For this illustration we are interested in the underlying true 
abundances underlying these two studies. Let  $\psi_1$ be 
the population size in 1995 and  $\psi_2$ be the size in 2001. 
Our main interest may lie in the annual growth rate underlying 
these two population sizes. We define 
$\rho=(\psi_2-\psi_1)/(6\psi_1)$, a simple 
(and in some sense approximate) 
definition of annual growth rate. 

\begin{figure}[h] 
\begin{center}
\includegraphics[scale=0.5]{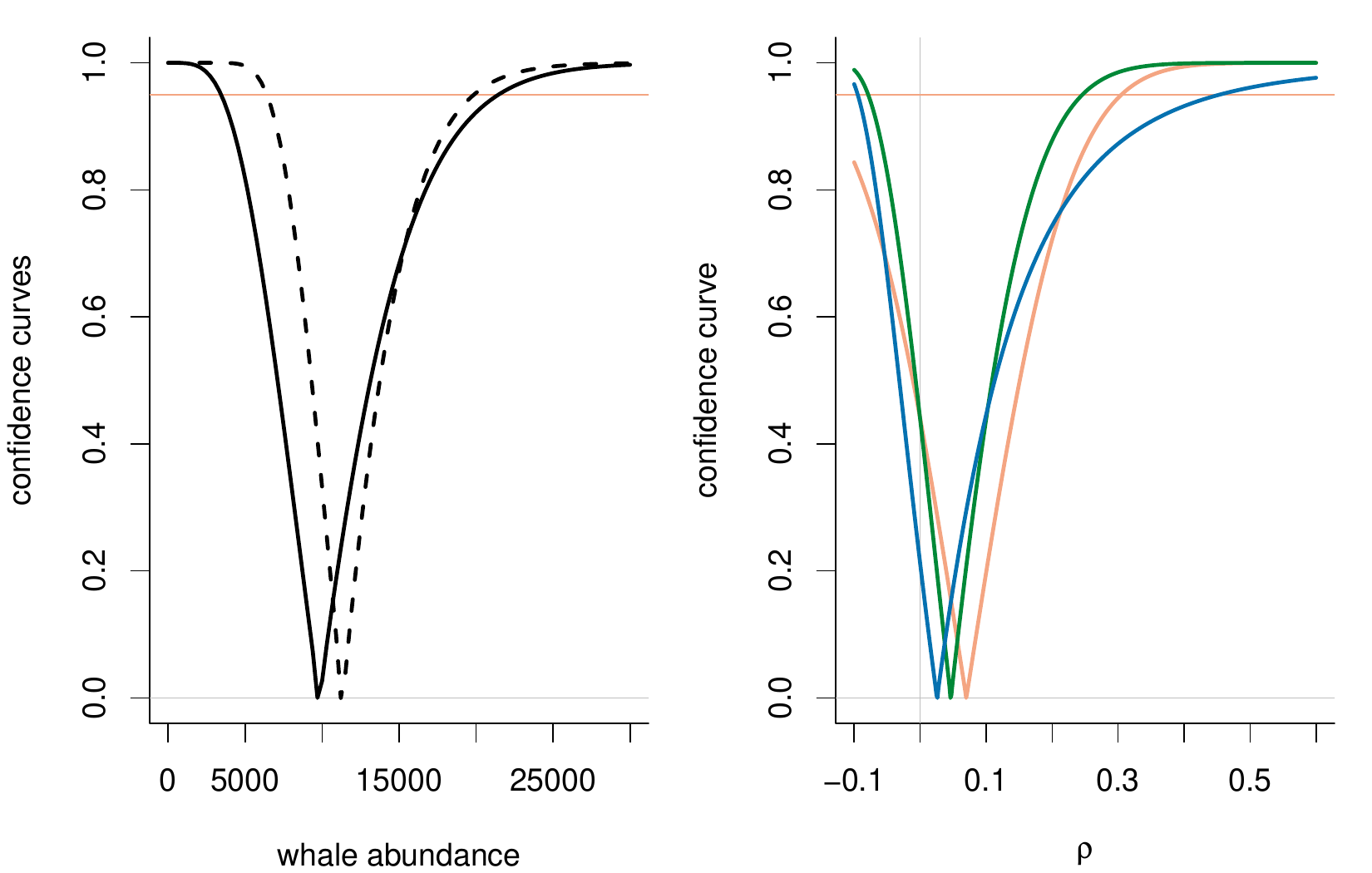}
\end{center}
\caption{Left panel: confidence curves for $\psi_1$ and $\psi_2$,
  the abundance of humpback whales in the North Atlantic
  in 1995 (fully drawn line) and 2001 (dashed line). 
Right panel: the confidence curve for 
$\rho=(\psi_2-\psi_1)/(6\psi_1)$ based on the two surveys (blue curve);  
the confidence curve based on prior information alone 
(orange curve); and the confidence curve combining 
the studies and the prior information (green curve).
See Section \ref{subsection:whales} and Table \ref{table:whales}.
\label{figure:whales} }
\end{figure}

The first step, {\it Independent Inspection}, 
requires us to construct CDs 
for $\psi_1$ and $\psi_2$ from the two surveys. 
In \citet[Ch.~10]{CLP}, certain methods are proposed 
and developed for constructing CDs based 
only on an estimate and a confidence interval. 
With a positive parameter, like abundance, one may use
\beqn
\hbox{II}\colon \qquad 
C(\psi_j,y)= \Phi\Bigl(\frac{h(\psi_j)-h(\hatt\psi_j)}{s}\Bigr)
\eeqn 
with a power transformation $h(\psi,a) = \sgn(a)\psi^a$; 
see also \citet{SchwederHjort13b} for some more discussion
of this approach (along with a different application,
essentially also using the II-CC-FF paradigm). 
In order to estimate the power $a$ and the scale 
$s$ the following two equations must be solved,  
\beqn
  \psi_L^a - \hatt\psi^a = -1.96\,s 
   \quad \hbox{\rm and} \quad \psi_R^a - \hatt\psi^a = 1.96\,s, 
\eeqn
where $[\psi_L,\psi_R]$ is the 95\% confidence interval
and $\hatt\psi$ the median confidence point estimate.   
For the whale abundance, we find 
$(a,s)$ equal to $(0.321,2.798)$ for 1995 and $(0.019,0.007)$ for 2001 
(a small value of $a$ indicates that the transformation
is nearly logarithmic).  
The corresponding confidence curves 
are shown in the left panel of Figure~\ref{figure:whales}. 
In this case the confidence log-likelihoods in 
the {\it Confidence Conversion} step are
easily obtained. For year $j$, 
\beqn
\hbox{CC}\colon\qquad 
\ell_{\con,j}(\psi_j) = -\half\{h_j(\psi_j)-h_j(\hatt\psi_j)\}^2/s_j^2. 
\eeqn
In the final {\it Focused Fusion} step, 
we sum the two confidence log-likelihoods, 
profile with respect to $\rho$,  
find the combined deviance function,  
and construct an approximative combined confidence 
curve by the Wilks theorem, as per Section \ref{section:CDs}: 
\beqn
\hbox{FF}\colon\quad 
\ell_{\fus,\prof}(\rho) &=& \max\{\ell_{\con,1}(\psi_1)
   +\ell_{\con,2}(\psi_2) \colon (\psi_2-\psi_1)/(6\psi_1)=\rho \}, \\
\cc^*(\rho)&=&\Gamma_1(2\{\ell_\fus(\hatt\rho)-\ell_\fus(\rho)\}).
\eeqn
Here we obtain the blue curve in the right panel of 
Figure~\ref{figure:whales}, with $\hatt\rho=0.026$ 
and a 95\% confidence interval $[-0.094,0.454]$. 

In some cases there may exist some expert knowledge 
pertaining to at least the focus parameter under study,
here the annual growth rate $\rho$, though not necessarily
for the full parameter vector of the combined models,
here $(\psi_1,\psi_2)$ the two population sizes. A proper Bayesian analysis
requires the statistician to have such a prior for 
$(\psi_1,\psi_2)$ -- without this ingredient,
there is no Bayes theorem leading to a posterior 
distribution for the model parameters, or indeed 
for $\rho$. The II-CC-FF scheme allows however 
incorporation of such partial prior information,
i.e.~a prior for $\rho$ without a prior for $(\psi_1,\psi_2)$.  
For this illustration we assume that whale biologists provide a normal
prior with expectation equal to $0.07$ and variance $0.12^2$.  
This prior may come from knowledge of other humpback 
whale populations or simulation-based life-history models 
(see for example \citet{zerbini2010}, 
giving a similar point estimate as we have used). 

The prior can be represented as a confidence curve, 
supplementing the confidence curve based on the two studies. 
In order to fuse the prior knowledge and the data 
we simply add the prior log-likelihood $\ell_B(\rho)$ 
to the confidence log-likelihoods, in the following way,
\beqn
\hbox{FF}\colon \qquad 
\ell_{\fus,\prof,B}(\rho) 
   &=& \max \{ \ell_{\con,1}(\psi_1)+\ell_{\con,2}(\psi_2)
   + \ell_B(\rho)\colon (\psi_2-\psi_1)/(6\psi_1)=\rho\} \\
   &=& \max_{\sigma_1} \{ \ell_{\con,1}(\psi_1)+\ell_{\con,2}(\psi_2)
      \colon (\psi_2-\psi_1)/(6\psi_1)=\rho\}+\ell_B(\rho) \\
   &=&\ell_{\fus,\prof}(\rho)+\ell_B(\rho). 
\eeqn 
We use `B' as subscript to indicate the in this instance
partial and perhaps lazy Bayesian, who does not give
a full prior for the model parameters, but contributes
a component, namely where it matters the most, about
the focus parameter. Of course the log-prior $\ell_B(\rho)$
employed here could have been obtained in the more 
careful and proper Bayesian way of having started with 
a full prior for $(\psi_1,\psi_2)$, and then 
a transformation, but we do suggest that expert knowledge
concerning focus parameters is more often put forward
directly, not via the full parameter vector in the 
fullest model. 

Importantly, this extended deviance function does still
have an approximate $\chi^2_1$ distribution, by the general
approximation arguments briefly discussed in
Section \ref{subsection:fusionlemmas}, 
unless the log-prior $\ell_B(\rho)$ is sharp and 
distinctly non-normal. One may conceptually and 
sometimes practically interpret the log-prior as having
resulted from real data in previous experiences,
in which case the $\ell_B(\rho)$ would be a genuine
profiled log-profile likelihood function from such a source. 
Also, as the sample sizes of the studies increase the 
information from the two studies will dominate the 
prior and we can safely continue to use the Wilks theorem. 
As expected, the confidence curve fusing the prior 
information and the information from the two studies 
lies between the original confidence curve and the 
prior confidence curve (see Figure~\ref{figure:whales}, right panel). 
It is also somewhat narrower than both.


\subsection{Olympic medians}
\label{subsection:quantiles}

\begin{figure}[h]
\centering
\includegraphics[scale=0.6]{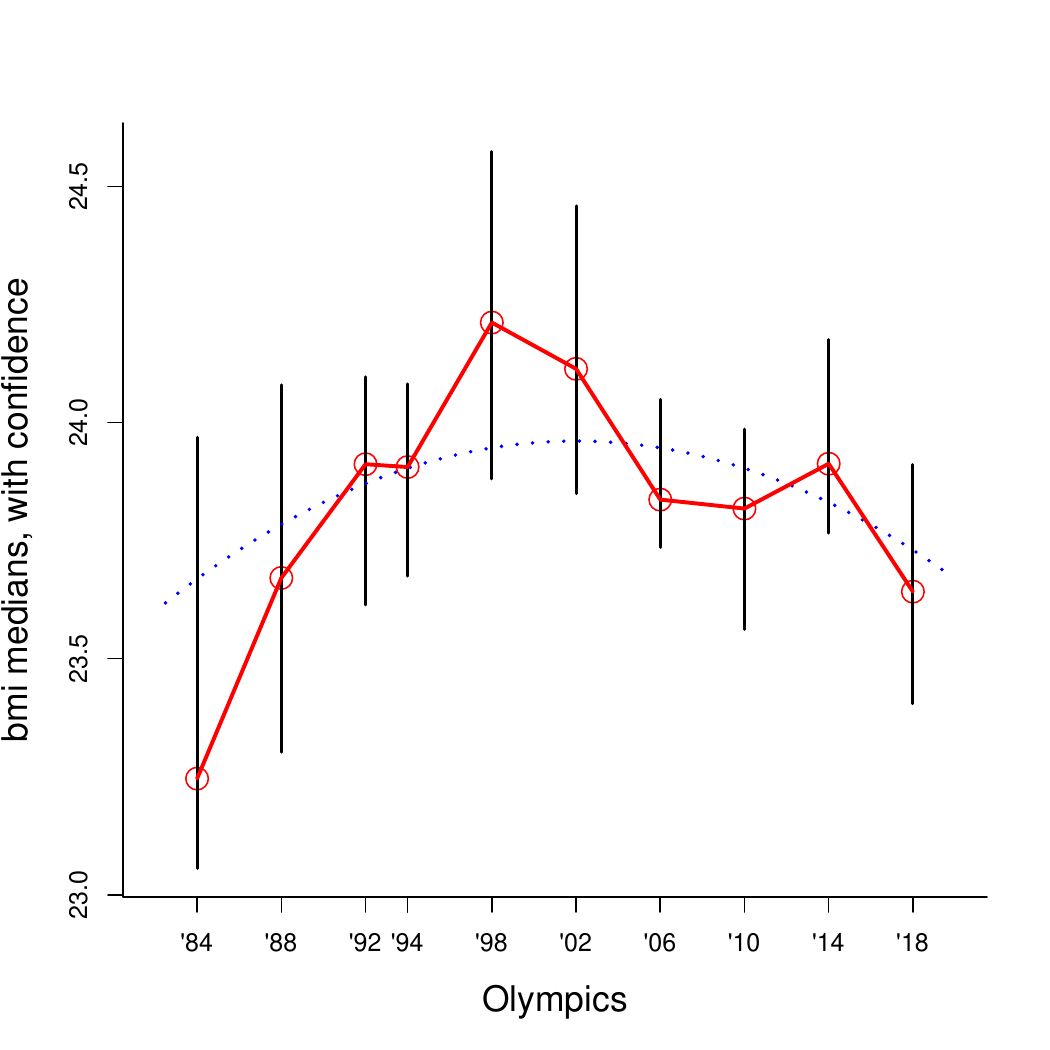}
\caption{Median BMI for men, ten last Olympics 1984 Sarajevo 
   to 2018 Sochi, with 90 percent confidence band,  
   and the fitted parabola (\ref{eq:parabel}).}
\label{figure:fig21}
\end{figure}

\noindent
This is an Olympic story about certain dynamic changes 
of the Body Mass Index distribution for speedskaters. 
We focus specifically on how the median of this
distribution has changed over time. Our use of the 
II-CC-FF scheme will involve first building highly
accurate nonparametric CDs for the medians, and 
then converting these to parametric log-likelihoods,
after which a dynamic model for the medians can 
be fused together in the end. Our methodology will
work with modest changes also for cases where 
interest lies in the evolution of any given quantile,
say the 0.90 quantile points of income distributions
over time across different strata. 

The BMI is defined as weight (in kg) 
divided by squared height (in metres). 
Figure \ref{figure:fig21} displays 
the median BMI, for the male participants, 
across the ten last Olympics 
(1984, 1988, 1992, 1994, 1998, 
2002, 2006, 2010, 2014, 2018, with the notable 
breaking of the Olympic rhythm from Albertville to Lillehammer),
marked as small red circles. The red line adjoining
these sample medians indicates that the BMI has
undergone a certain evolution, with a marked 
increase up to perhaps 1998 Nagano or 2002 Salt Lake City,
followed by a downward trend. Even though these
changes perhaps do not qualify as drastic, 
and athletes with 24.5 do not differ very much 
from those with 23.5, they are nevertheless 
interesting enough to be discussed in the 
proper fora. Potential underlying reasons 
behind such an evolution, with an up and a down 
over ten Olympiads, include 
(i) the prospect of doping, with building of 
extra muscle power, etc., and the alleged 
cleaning of the sport; 
(ii) the klapskate and the higher speed, which
arguable favours technical skills over pure power; and
(iii) the introduction of the team event 
and mass starts, which points to statistical
discrepancies in the populations of male Olympic 
skaters from e.g.~2002 and 2018. 

\begin{figure}[h]
\begin{center}
  \includegraphics[scale=0.4]{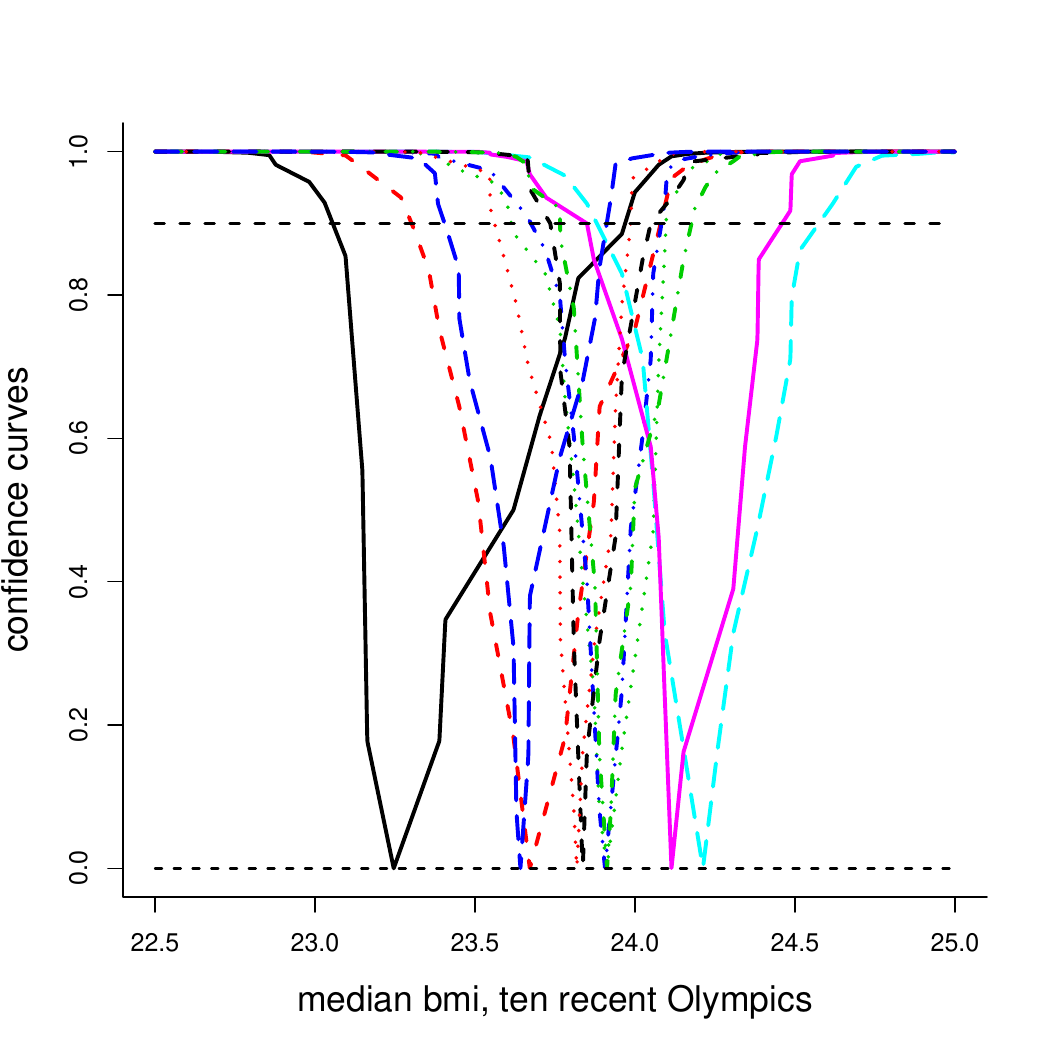}
  \includegraphics[scale=0.4]{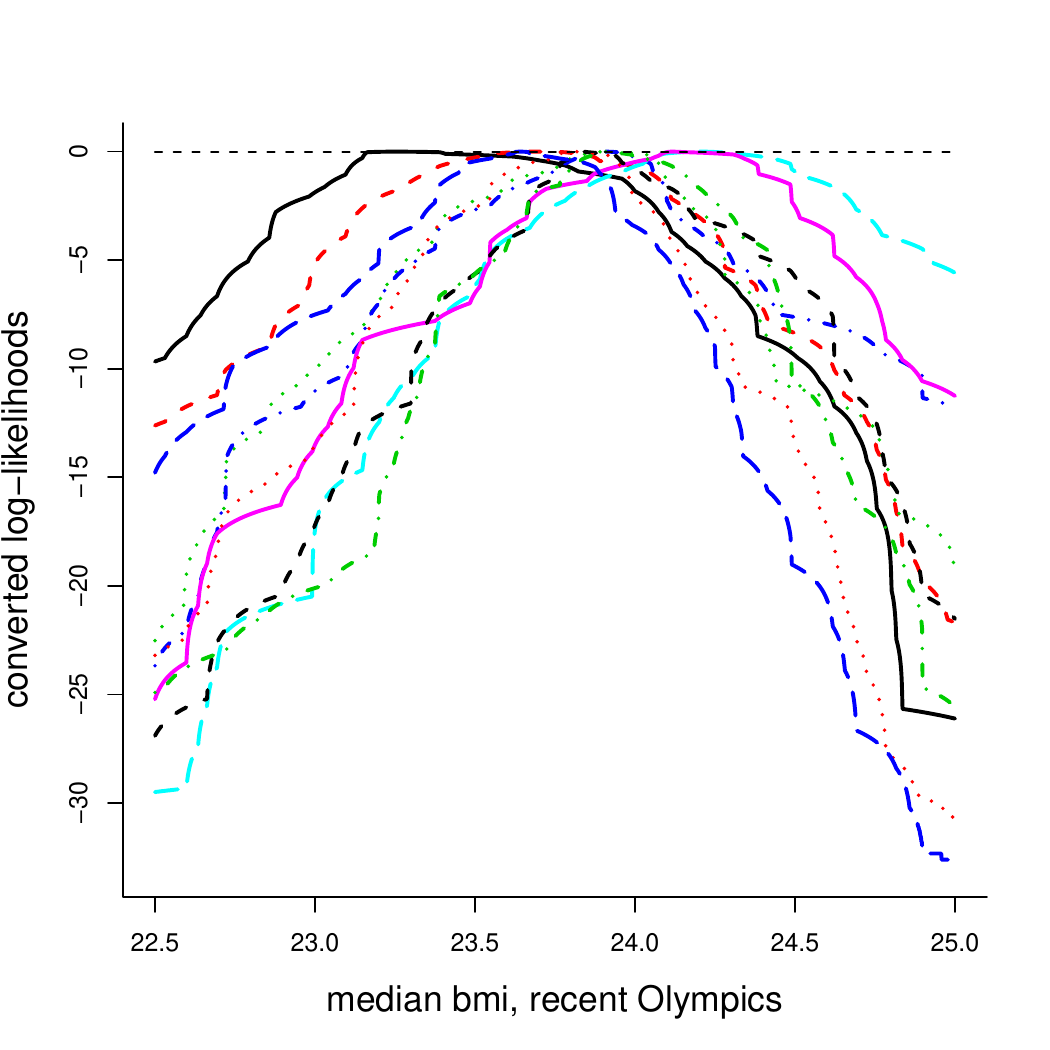}
\end{center}
\caption{Left: Nonparametric confidence curves $\cc_j(\mu_j)$ 
   for the medians of the BMI distributions, 
   for the ten last Olympics 1984 to 2018. 90 percent
   confidence intervals are read off from the 
   horizontal dashed line at 0.90. 
   Right: Converted log-likelihood contributions 
   $\ell_{\con,j}(\mu_j)$ for the ten Olympic medians.}
\label{figure:fig2223} 
\end{figure}

These themes motivate the following investigation, 
to look both for significant changes and for the position 
of a potential top-point for the evolution of the BMI
distribution over time. Let $\mu_j$ be the population
median at Olympics $j$, for occasions $j=1,\ldots,k$. 
We first need CDs for these, constructed nonparametrically.
With $y_{j,(1)}<\cdots<y_{j,(n_j)}$ the ordered sample 
from Olympics $j$, assumed to come from a continuous 
distribution with positive density $f_j$ and cumulative $F_j$, 
we start from the exact calculation 
\beqn
\Pr_{f_j}\{ \mu_j\le y_{j,(r)} \}
=\Pr\{F_j(\mu_j)\le F_j(y_{j,(r)})\}
=\Pr\{\half\le U_{j,(r)}\}, 
\eeqn  
with $U_{j,(1)}<\cdots<U_{j,(n_j)}$ an ordered i.i.d.~sample
from the uniform distribution on the unit interval. 
But these have known Beta distributions.
We therefore define the $C_j(\mu_j)$ 
for all values of $\mu_j$ by first setting 
\beqn
C_j(y_{j,(r)})=1-\Be(\half,r,n_j-r+1)
\quad {\rm for\ }r=1,\ldots,n_j, 
\eeqn 
featuring cumulative Beta distribution functions,
and then applying linear interpolation between
the ordered data points.
The full confidence curves $\cc_j(\mu_j)=|1-2\,C_j(\mu_j)|$ 
are displayed in Figure \ref{figure:fig2223} (left panel) 
and they are then converted to log-likelihood components 
$\ell_{\con,j}(\mu_j)=-\half\Gamma_1^{-1}(\cc_j(\mu_j))$ 
(right panel). 
The 90 percent confidence intervals for the 
ten medians shown in Figure \ref{figure:fig21} 
are computed from the $\cc_j(\mu_j)$. 
Note that these confidence curves and intervals
are fully nonparametric. They can be shown 
to be highly accurate, even for smaller sample sizes,
and work better than alternative methods 
involving approximate normality with 
estimates of standard errors. The resulting intervals 
deviate from symmetry, reflecting underlying 
asymmetry in the distribution of the BMI.  

Next we model the medians $\mu_j$ dynamically as 
\beq
\label{eq:parabel}
\mu_j=\beta_0+\beta_1x_j+\beta_2x_j^2 
   \quad {\rm for\ }j=1,\ldots,k, 
\eeq 
where $x_j=t_j-t_1$ is the time passed 
since the first of the Olympics to no.~$j$. 
Such a model is found to be entirely adequate 
via AIC analysis. The parameters of this parabola 
will be such that it has a maximum point 
\beq
\label{eq:xstar}
x^*=x^*(\beta_0,\beta_1,\beta_2)=-\beta_1/(2\beta_2)
\eeq 
within the range from the first to the last of these Olympics, 
see again Figure \ref{figure:fig21}. 
Now the FF step of our general recipe leads to 
$\ell_\fus(\beta_0,\beta_1,\beta_2)
=-\half\sumjk \Gamma_1^{-1}(\cc_j(\beta_0+\beta_1x_j+\beta_2x_j^2))$, 
and then to the profiled version 
\beqn
\ell_{\fus}(x^*)
=\max\{\ell_{\fus}(\beta_0,\beta_1,\beta_2)\colon 
  (\beta_0,\beta_1,\beta_2){\rm\ fitting\ with\ }x^*\}.
\eeqn 
Using the FF step of our II-CC-FF is also 
equivalent to working with the fused deviance function
$D_\fus(x^*)=2\{\ell_\fus(\hatt x^*)-\ell_\fus(x^*)\}$,
with $\hatt x^*$ the appropriate function of
the overall maximisers of $\ell_\fus(\beta_0,\beta_1,\beta_2)$.

\begin{figure}[h]
\centering
\includegraphics[scale=0.6]{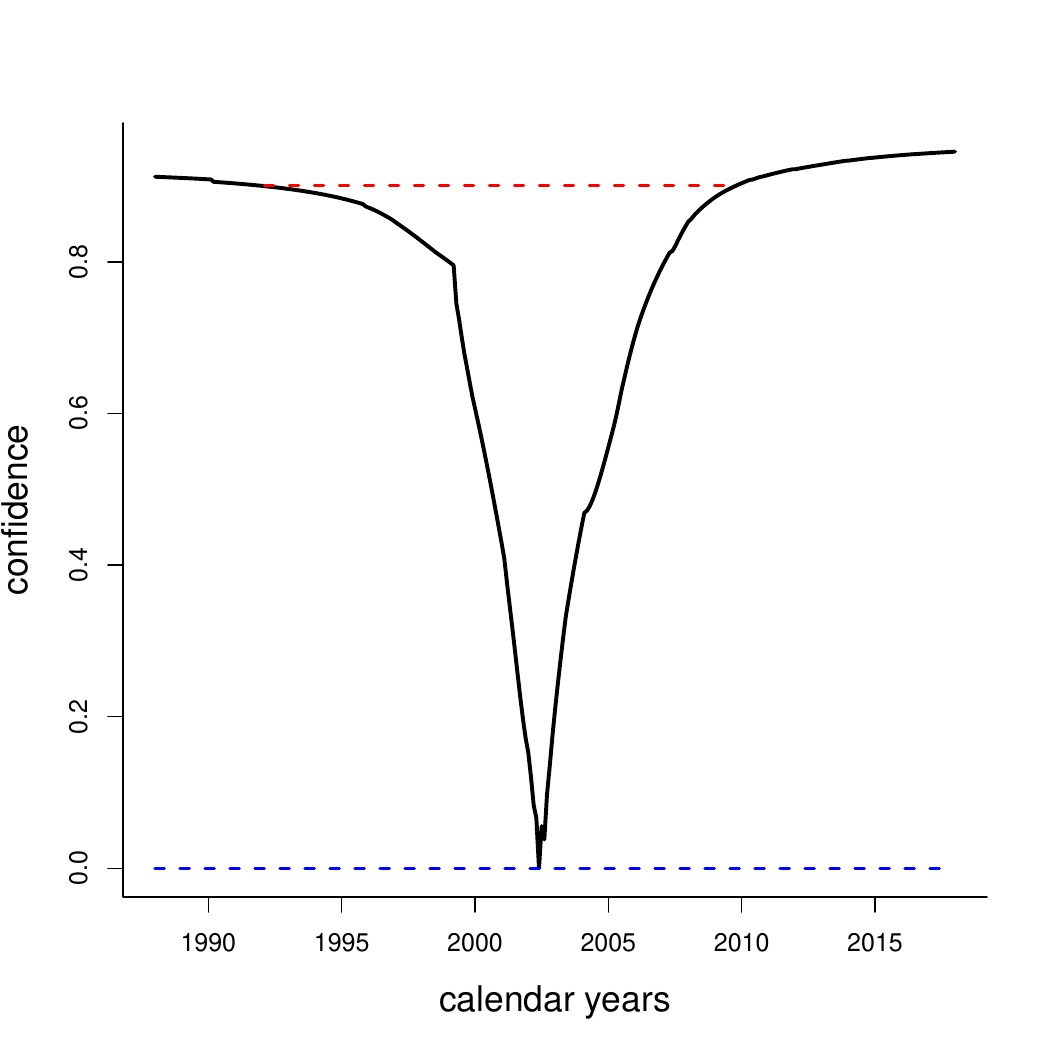}
\caption{Focused Fusion confidence curve for 
   $x^*=-\beta_1/(2\beta_2)$, the maximum point for 
   the parabola in the model for Olympic BMI medians.}
\label{figure:fig24}
\end{figure}

Carrying out this machinery leads first to 
the fitted parabola shown as the blue dashed 
curve in Figure \ref{figure:fig21}. 
The linear and quadratic coefficients $\beta_1$ and $\beta_2$  
are very significantly positive and negative,
respectively, as shown via Wald ratios; 
also, the estimated top-point is at $\hatt x^*=2002.4$. 
Secondly, using the distribution approximation 
$D_\fus(x^*)\sim\chi^2_1$ we can execute the FF step
and compute the confidence curve 
$\cc_\fus(x^*)=\Gamma_1(D_\fus(x^*))$. 
It is displayed in Figure \ref{figure:fig24}, 
with the partial ruggedness and asymmetry reflecting 
the nonsmoothness of the ten nonparametric 
sample median operations. We have been able 
to use confidence conversion, from 
nonparametric confidence curves to log-likelihood 
contributions and then back again to a focused fusion
confidence curve. The 90 percent confidence interval 
for the point of maximum median BMI for male 
speedskaters is from 1992.1 to 2009.7. 

\subsection{The long peace} 
\label{subsection:pinker}

Our last illustration concerns the use of the II-CC-FF 
framework in a highly non-standard setting, where one wishes 
to combine hard data, sources that inform directly on the focus 
parameter, with softer data sources, which only contain indirect 
or noisier information about the focus parameter. 
This kind of combination has wide potential in various 
fields where `soft' data could be based on webscraping, 
using twitter accounts or other social media, 
but raises specific issues and challenges. 
The question we investigate here is the extent of statistical 
evidence for The long peace, the period of relative peace 
and stability following the second world war (and still lasting, 
presumably). Specifically, do we find evidence of a change-point 
$\tau$ when analysing the sequence of battle deaths in 
interstate wars between 1823 and today? 

There are many components, issues, and details involved
in this application story, and a fuller version
is reported on in Appendix \ref{section:appendixC}.
Here we outline the main statistical ingredients. 
First, the question has been investigated in 
\citet*{CunenHjortNygard2020} using the Correlates of War (CoW) 
dataset \citep{SarkeesWayman10}. 
The authors found evidence of an abrupt change 
in the battle death distribution at some point 
after the second world war,  
from a distribution with a high median battle death 
to a distribution with a lower median (and also a less heavy tail). 
This involved establishing a certain three-parameter model
for the battle deaths distribution, with parameters
changing at time $\tau$, itself an unknown change-point parameter. 
We may view the relevant statistical information 
as represented by an FF based log-likelihood
contribution $\ell_{B,\prof}(\tau)$. 
Here the aim is to extend this analysis 
and investigate whether there might be benefits in 
combining the battle death data with other sources 
assumed to be informing on $\tau$.

Some political scientists consider the aforementioned decrease 
in battle deaths to reflect a moral and political shift 
within a large portion of the world's population. At some point 
in the 20th century, it is argued, 
the perception of war changed, from 
being seen as something natural and inevitable, 
sometimes even positive, to being perceived as highly negative, 
evil and unacceptable; cf.~\citet[Ch.~5]{Pinker11}.  
This change in norms has likely manifested itself in various 
ways, including cultural, artistic and political expressions, 
for example through text. We have therefore collected 
sequences representing the usage of certain relevant words
or phrases, like `anti-war' or `pacifist', 
and then attempted to combine the change-point inference 
from such an Ngram analysis (suggested to us by Steven Pinker, 
personal communication), with the change-point inference 
from the battle deaths data. Such statistical work,
along with data collection and separate modelling efforts, 
leads to an overall log-likelihood contribution 
$\ell_{N,\prof}(\tau)$ for the potential change-point $\tau$.  
These combination efforts lead to
an overall log-likelihood fusion function 
\beqn
\ell_\fus(\tau)=w_B\ell_{B,\prof}(\tau)+w_N\ell_{N,\prof}(\tau), 
\eeqn 
with relative importance weights $w_B$ and $w_N$,
involving a separate discussion. Other applications,
involving the combination of perhaps very different 
data sources, would similarly involve separate 
and perhaps partly subjective analyses 
for deciding on such relative importance weight parameters. 
For the present application, with battle deaths 
the hard data and Ngrams the soft data, 
more details are presented in Appendix \ref{section:appendixC}, 
along with our tentative concluding confidence curve $\cc^*(\tau)$. 

\bigskip
{\bf Acknowledgements.} 
The work reported on here has been partially funded 
via the Norwegian Research Council's project
{\it FocuStat: Focused Statistical Inference With Complex Data},
led by Hjort. The authors are grateful to comments from
Steven Pinker, and to Tore Schweder and Emil Stoltenberg 
for always fruitful discussions related to issues
and methods worked with in this paper. 
Constructive comments and suggestions from anonymous
reviewers and an associate editor have contributed
to a better paper. 
The data on heights and weights and hence BMI 
for Olympic speedskaters stem from files collected
over the years by N.L.H., and in this endeavour
he has been helped by fellow speedskating historians
Arild Gjerde and Jeroen Heijmans. 



{{\footnotesize 
\bibliographystyle{biometrika}
\bibliography{iiccff_bibliography8.bib}
}}

\bigskip\noindent 
Below follow extra material, organised as
Appendixes A, B, C. 
The first gives more details pertaining to
our simulations, summarised in Section \ref{section:performance}.
Appendix B reports on our study of
how to combine hard data (battle death counts,
for 95 major wars) and soft data (Ngram monitoring 
studies for certain key-words, sampled from 
tens of millions of books). Finally details for 
how the II-CC-FF deals with the Neyman--Scott 
situation are in Appendix C.   

\appendix

\section{Additional details on the simulations}
\label{section:appendixB}

Here we provide some more details on the simulations presented in 
Sections \ref{subsection:basicREsim} and 
\ref{subsection:meta2x2sim}. In particular, we provide the {\tt R} 
function calls for the methods with which we have compared our II-
CC-FF methods. 
For Section \ref{subsection:meta2x2sim} we also present some 
additional results which were described in the text.

\subsection{The basic random effect model}

For the Hartung-Knapp-Sidik-Jonkman method we used the 
implementation in the {\tt metafor} package, and the following 
function call
\begin{verbatim}
rma.uni(yi= sumM[ ,1], sei= sumM[,2], test="knha")
\end{verbatim} 
The matrix {\tt sumM} provides the $k$ estimates for $\psi_j$ in 
the first column, and estimates for $\sigma_j$ in the second.
With this function, the default estimator for $\tau$ is the REML 
estimator. 
Note that there is a small probability that this function fails if the 
algorithm for finding the REML estimator does not converge, and in 
those (very few) cases we used the DerSimonian-Laird method 
instead.

For the CD combination method of \citet{XieSingh2011}, we used 
the implementation in the {\tt gmeta} package,
\begin{verbatim}
gmeta(sumM[ ,1:2], method="random-reml", gmo.xgrid=psival)
\end{verbatim}
The matrix {\tt sumM} provides the $k$ estimates for $\psi_j$ in 
the first column, and estimates for $\sigma_j$ in the second. The 
vector {\tt psival} is a grid of 500 values for $\psi_0$ (but with 
limits adjusted according to $k$ and $\tau$). This was the same 
grid we used for our II-CC-FF methods.
With this function, the default estimator for $\tau$ is the REML 
estimator. 
Note that there is a somewhat significant probability that this 
function fails if the algorithm for finding the REML estimator does 
not converge, and in those cases we used the DerSimonian-Laird 
method instead.
The option {\tt method="random-reml"} is actually one of six 
choices for the basic random effect model which are implemented 
in the {\tt gmeta} package. We did not find any recommendations 
concerning which method to choose, except that we avoided 
the methods denoted as {\tt "robust"}, 
since these are meant for situations where there are outlying 
studies, according to \citet{XieSingh2011} 
(and which is not the case in our set-up).

\subsection{Meta-analysis of $2 \times 2$ tables}

For the fixed effect meta-analysis we used the Mantel-Haenszel 
method which we implemented ourselves based on the formulas 
given in \citet{piaget2019}. For the variance of the estimator of the 
risk difference (equation (19) in \citet{piaget2019}) there seemed to 
be a small typo, and we replaced the minus sign in the middle with 
a plus sign. 

Note that for the odds ratio and risk ratio, the 
Mantel-Haenszel method produces indefinite estimates when 
there is not a single event in the entire control arm of a study 
(i.e. all $y_{0,j}$ are equal to zero). The {\tt gmeta}
also fails in this situation, which happens 
occasionally in the set-up shown in Figure \ref{figure:sim1}: 
in 7\% of the rounds 
for $k=5$ for instance. For the sake of fair comparisons we 
have removed these rounds from all the methods. Further, the 
Mantel-Haenszel produces indefinite variance estimates also when 
there is not a single event in the entire treatment arm of a study 
(i.e. all $y_{1,j}$ are equal to zero). This situation happens often in 
our set-up (around 50\% of the time for $k=5$) and in these cases 
we defined the 95\% confidence interval from the Mantel-Haenszel 
method as spanning the entire real line. 
The II-CC-FF methods never fail/crash: in situations with zero 
events in the entire control arm or the entire treatment arm, the 
resulting confidence curves for the log odds ratio 
(or the log risk ratio) will have a point-mass in infinity or minus 
infinity, respectively. 

We used the {\tt gmeta} package for the odds ratio and the risk 
difference. With the following function calls,
\begin{verbatim}
gmeta(dd,gmi.type="2x2", method="exact1")
gmeta(dd,gmi.type="2x2", method="exact2")
\end{verbatim}
The matrix {\tt dd} has $k$ rows and four columns. The first and 
third column are the number of events in treatment and control 
group respectively. The second and fourth column are the sample 
sizes in each group. The options {\tt "exact1"} and {\tt "exact2"} are 
chosen according to the {\tt gmeta} documentation.

For the risk ratio we used the {\tt exactmeta} package with the 
following function call,
\begin{verbatim}
meta.exact(dd[ ,c("y0","y1","m0","m1")], type="rate ratio", print=F)
\end{verbatim}
Note here that we used the  {\tt "rate ratio"} option which actually 
computes the ratio of two Poisson rates. There exists an option 
{\tt "risk ratio"} in the package, but that option was prohibitively 
slow. When event probabilities are small the Poisson rates 
and binomial proportions will be close to equal. If anything 
we are giving the {\tt exactmeta} package an edge compared to 
the other methods here. 

\begin{figure}[h]
\centering
\includegraphics[scale=0.55]{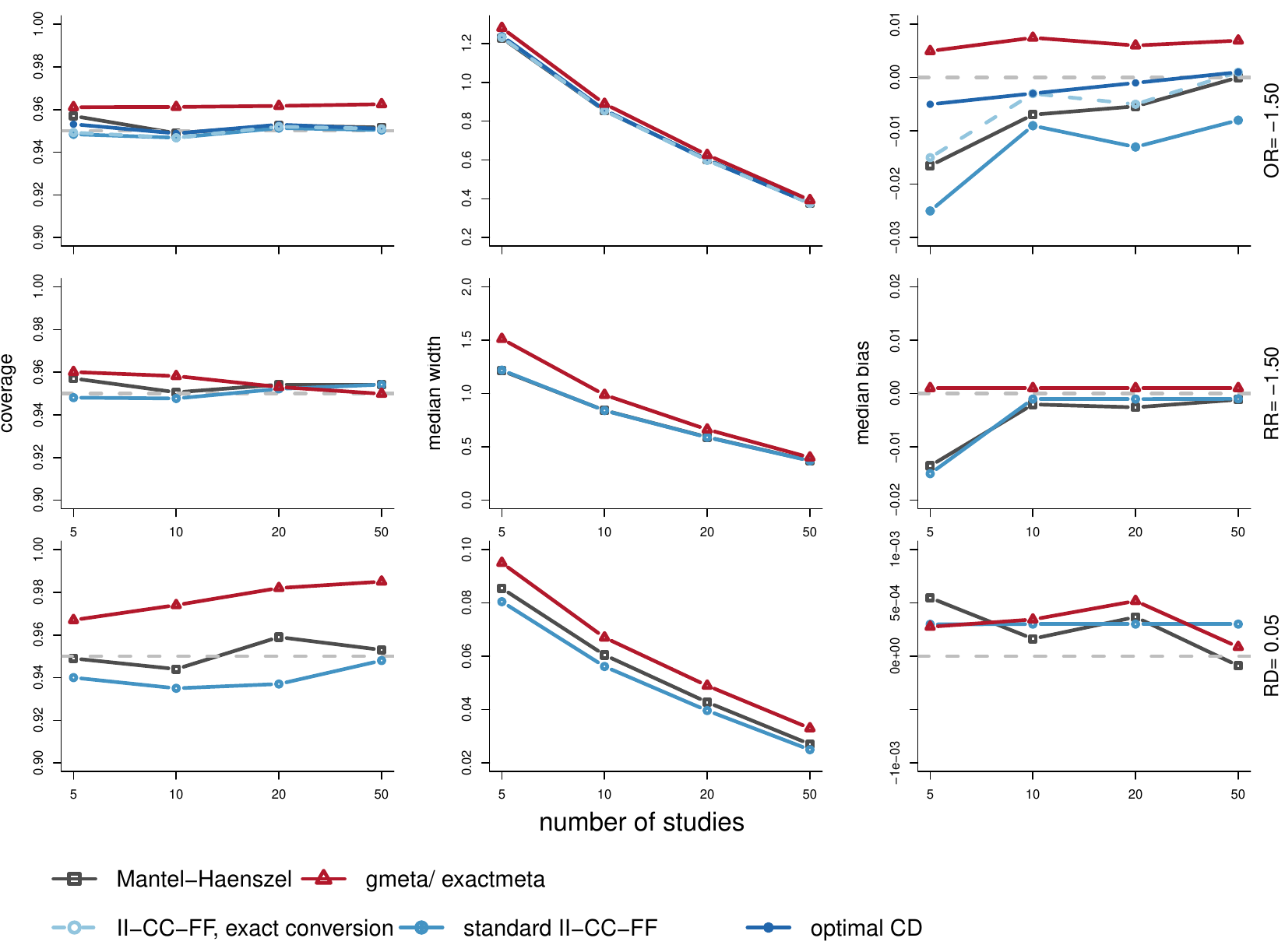}
\caption{Simulation results for fixed effect meta-analysis of 
$2 \times 2$ tables in a setting with not particularly rare events. 
The left column gives the realised coverage rate of 95\% confidence 
intervals, the middle column gives the median width of these 
intervals and the right column gives the median bias of the point 
estimate coming from each of the methods. The top row gives the 
results for the (log) odds ratio, the middle row for the (log) risk 
ratio and the bottom row gives the results for the risk difference.}
\label{figure:sim1app}
\end{figure}

As mentioned in Section \ref{subsection:meta2x2sim} we 
have run similar simulations as the ones shown there, 
but in a setting with less rare events. These results 
are shown in Figure \ref{figure:sim1app}. There we have 
median event probability in the control group is 0.1, 
instead of 0.005 in the main text.

Here we also provide two additional notes on 
the methods we have used and the links between them. (1) 
For the odds ratio, the \citet{liu2014}  method, the optimal CD 
method and the II-CC-FF method with exact conversion are all 
three closely related to the Fisher exact test \citep{fisher1954}. 
Their performance is very different, however, and that is due 
to different strategies for the combination of the information 
from each source. 
(2) \citet{piaget2019} report that the estimators from the Mantel-
Haenszel procedure correspond to their respective maximum 
likelihood estimators when the sample sizes in the two groups have 
a constant ratio for all studies. We have simulated data where we 
allow for some variability in the ratio between the sample sizes, but 
this might still constitute a partial explanation for why our II-CC-FF 
methods have relatively similar behaviour as the Mantel-Haenszel 
method.

For the random effects simulations, we compared our methods with 
the modified Simmonds-Higgins method. The method is 
implemented in the {\tt metafor} package, 
and we used the following function call
\begin{verbatim}
rma.glmm(measure="OR",ai=y1sim,bi=m1s-y1sim,ci=y0sim,di=m0s-y0sim,model="UM.FS")
\end{verbatim}
The vectors {\tt y1sim}, {\tt m1s-y1sim}, {\tt y0sim}, 
and {\tt m0s-y0sim} are all $k$-dimensional and 
represent the number of events and non-events in the treatment 
and control group, respectively. The option {\tt "UM.FS"} 
denotes the modified Simmonds-Higgins method.

\begin{figure}[h]
\centering
\includegraphics[scale=0.45]{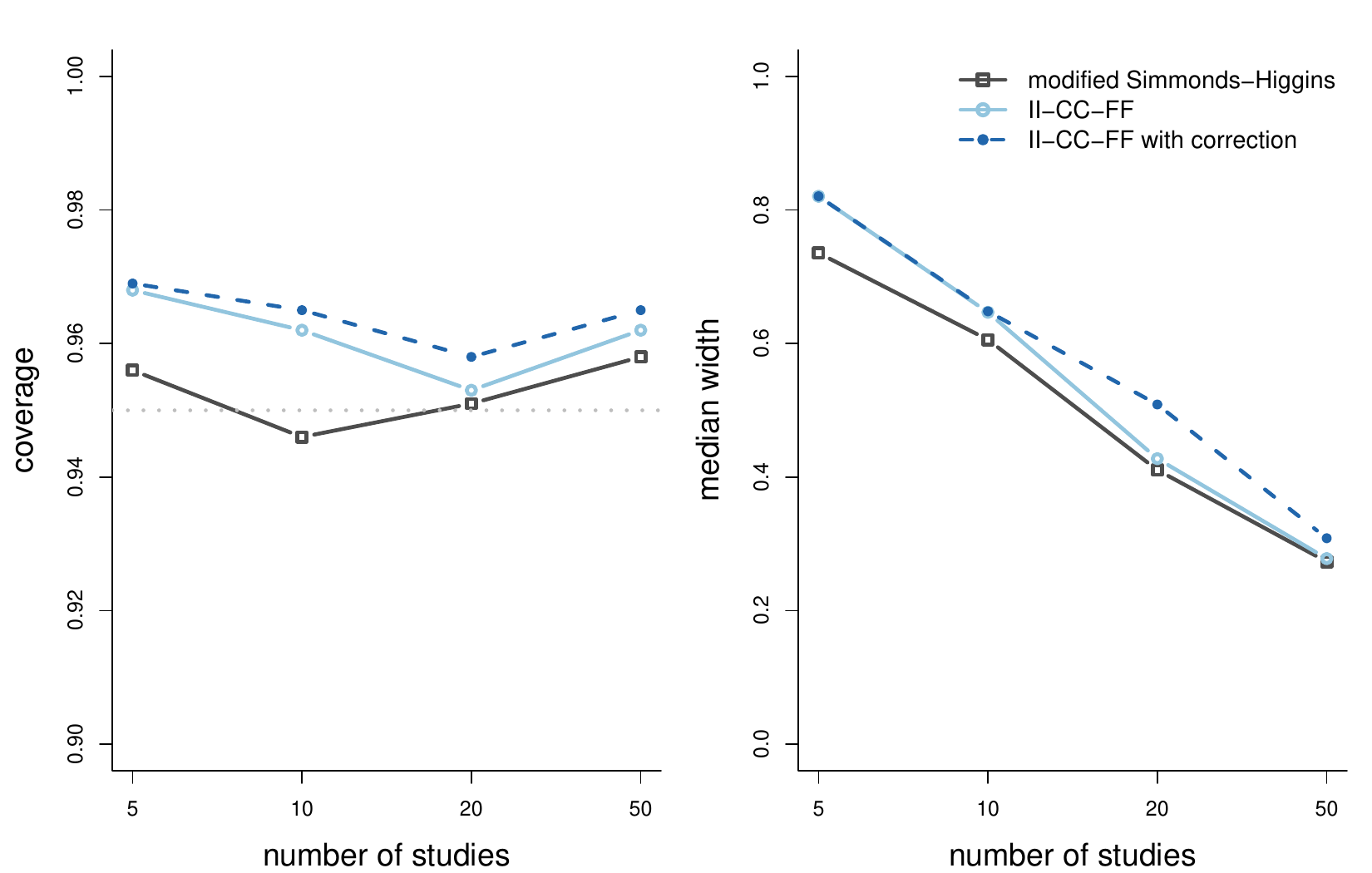}
\caption{Simulation results for random effect meta-analysis of 
$2 \times 2$ tables with little heterogeneity in the treatment 
effects. The left plot gives the realised coverage rate of 95\% 
confidence intervals, the right plot gives the median width of these 
intervals. }
\label{figure:sim2app}
\end{figure}

As mentioned in Section \ref{subsection:meta2x2sim} we have run 
similar simulations as the ones shown there, but in a setting 
with little heterogeneity in the treatment effects. 
These results are shown in Figure \ref{figure:sim2app}. 
There we have used $\tau=\sqrt{0.024}$ 
(instead of $\tau=\sqrt{0.168}$ in the main text). 
We have furthermore also increased the sample sizes, 
with $m_{1,j} \sim \unif(10,100)$. 

\section{Combining hard and soft data:
     battle deaths and Ngrams}
\label{section:appendixC}

Here we provide more details for Application \ref{subsection:pinker}.
The grand but imprecisely posed statistical meta-question
to be addressed is whether the world is experiencing
something worthy of being called {\it The long peace}, 
see \citet{Gaddis89} and the long chapter in
\citet{Pinker11} for extensive discussions; 
whether a change-point from `before' to `after'
may be identified; and in that case with what precision.
The `hard data' to be used are from data source (B), say, 
the series of battle deaths for the 95 major interstate wars,
from 1823 to the present, with at least 1000 deaths.
This is to be combined with `soft data' from data source (N), 
what we might find via the Google Books Ngram Viewer
machinery \citep{Michel2010}, monitoring the changes and evolution
over time regarding the frequency with which certain words
or key phrases are used in tens of millions of books,
with corpora currently spanning the era from 1800 to 2010.

These questions have already been addressed in
\citet*{CunenHjortNygard2020}, using then only the hard data (B).
They developed the three-parameter model
with cumulative distribution function 
\beqn
F(z,\mu,\alpha,\theta)
=\Bigl[{\{(z-1000)/\mu\}^\theta \over 1+\{(z-1000)/\mu\}^\theta}\Bigr]^\alpha
\quad {\rm for\ }z\ge1000,
\eeqn
which has the type of heavy tails seen
in various analyses of such violence data. 
It was demonstrated first that the battle deaths
series has not been constant over time,
and then that there is an identifiable change-point $\tau$,
with one parameter vector before and another
parameter vector after $\tau$.
Their point estimate is 1950, the Korean war,
and the authors constructed a full confidence curve,
say $\cc_B(\tau)$, for the change-point, using
methodology developed in \citet*{CHH18}.
These methods do employ log-likelihood profiling,
so there is indeed an $\ell_{B,\prof}(\tau)$
involved, though distributions of deviance functions
are far from chi-squared; also, the resulting
confidence sets might be unions of disjoint
intervals. This aspect of such confidence methods
for change-points is seen in Figure \ref{figure:ccBN}. 
The curve reveals a point estimate for the change-point
in 1950, but with considerable uncertainty; 
the years 1939 and 1965 are also considered likely 
candidates for the change.

We will now discuss how to arrive at
a supplementary $\ell_{N,\prof}(\tau)$
using Ngram counts, which warrants a separate
brief detour in our statistical discussion.  
Some political scientists consider the aforementioned decrease 
in battle deaths to reflect a moral and political shift 
within a large portion of the world's population. At some point 
in the 20th century, it is argued, 
the perception of war changed, from 
being seen as something natural and inevitable, 
sometimes even positive, to being perceived as highly negative, 
evil, and unacceptable; cf.~\citet[Ch.~5]{Pinker11}.  
This change in norms has likely manifested itself in various 
ways, including cultural, artistic and political expressions, 
for example through text. We will therefore collect sequences 
representing the usage of certain relevant words, 
and then attempt to combine the change-point inference 
from such an Ngram analysis (suggested to us by Steven Pinker, 
personal communication), with the change-point inference 
from the battle deaths data. 
 
\begin{figure}[h] 
\begin{center}
\includegraphics[scale=0.5]{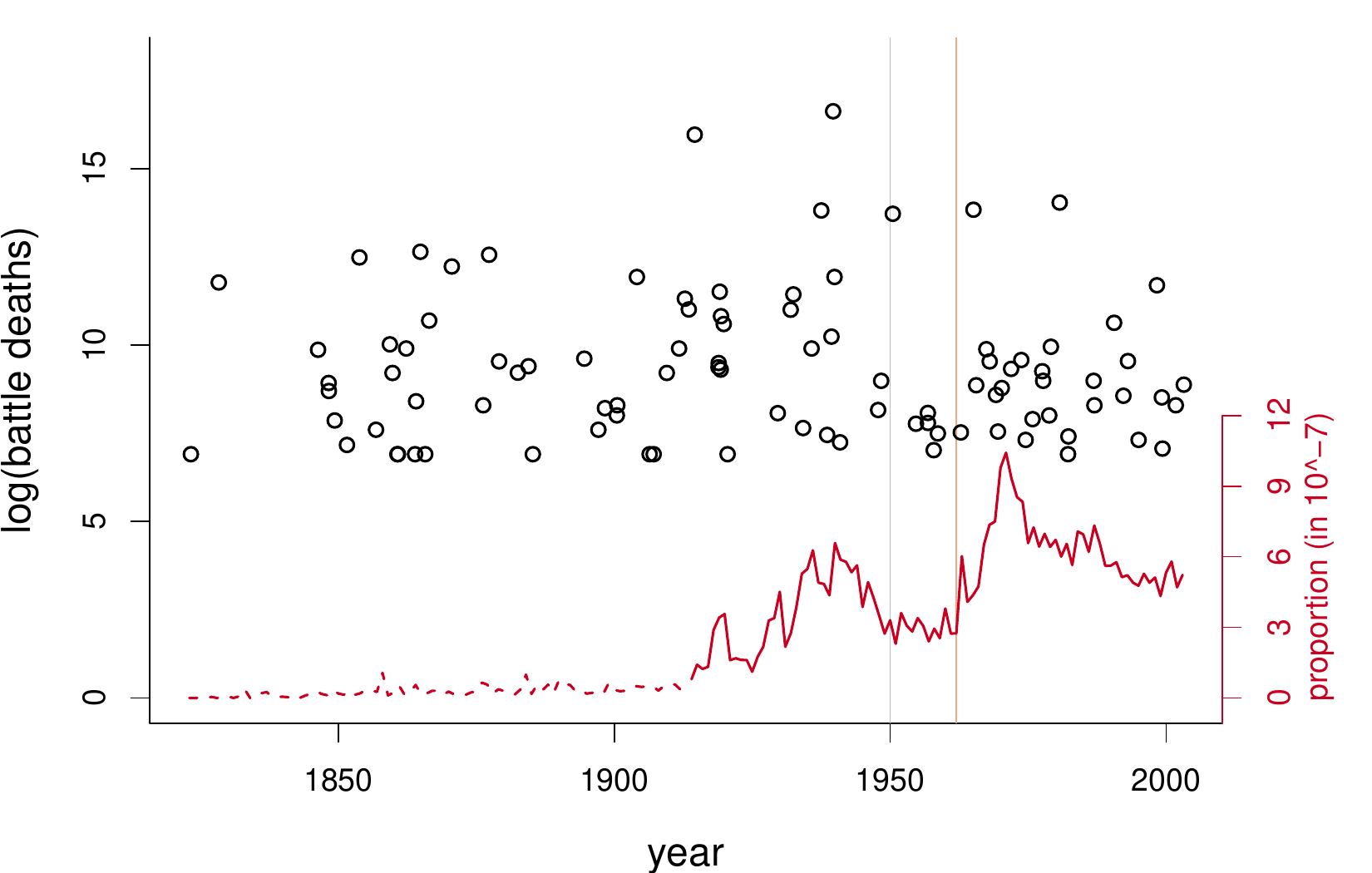}
\end{center}
\caption{The points represent the battle deaths, on log scale, 
for 95 wars between 1823 and 2003. Note that the CoW dataset 
only includes wars with at least 1000 battle deaths. The vertical 
grey line gives the point estimate for the change-point based 
on the battle deaths data. The red line shows the Ngram for `anti-war', 
i.e.~the number of times that word appears in each year divided 
by the total number of words in the corpus from each year. 
The counts between 1823 and 1913 were not used in the change-point 
analysis (and hence dashed). The vertical red line gives 
the point estimate for the change-point based on the Ngram.
\label{figure:data} }
\end{figure}

\begin{figure}[h] 
\begin{center}
\includegraphics[scale=0.5]{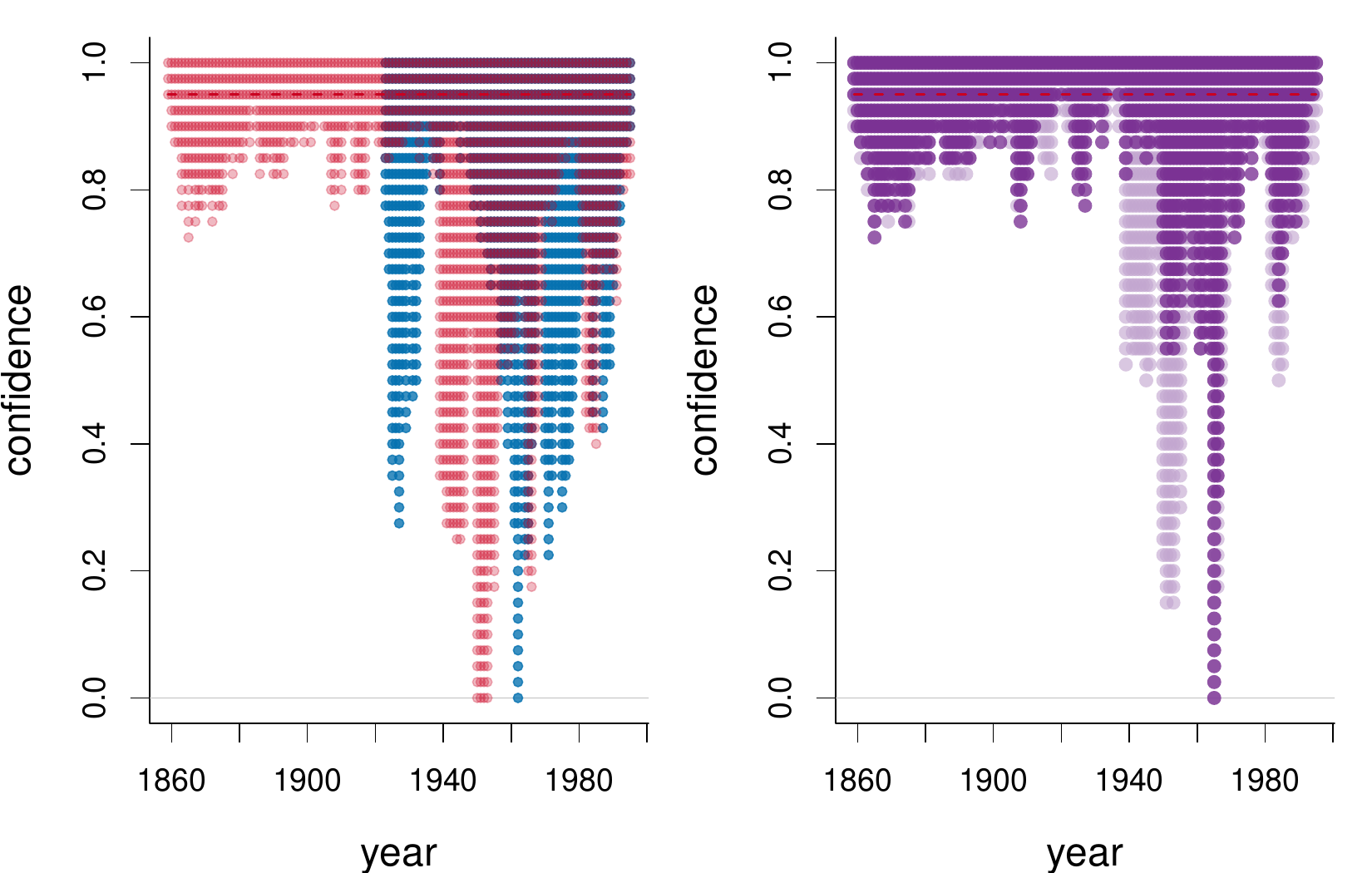}
\end{center}
\caption{Left panel: in red the confidence curve based 
on the battle death data (point estimate 1950), 
in blue the one based on the Ngram (point estimate 1962). 
Right panel: combined confidence curves, dark violet 
with no down-weighting, light violet with down-weighting 
of the Ngram information. Both these two curves 
give a point estimate equal to 1965.
\label{figure:ccBN} }
\end{figure}

We aim at providing a more thorough analysis of these 
questions in future work, also factoring in yet other
information sources; `signals for war' indexes of the type
worked with in \citet{Chadefaux14}, perhaps assisted 
by machine learning algorithms to sample news sources
from around the world and reading their pre-war-signals,
would be worthwhile to include. 
For the sake of the present illustration we limit ourselves 
to Ngram analysis of one word, however: `anti-war'. 
We collected the rate of usage of `anti-war' for each 
year between 1823 and 2003 from the Ngram viewer, 
see Figure \ref{figure:data}. 
The rate of usage is the number of times that word appears 
in each year divided by the total number of words in the 
Google Books corpus from each year. For a more thorough 
analysis we would build a score based on several such
Ngrams, or even a joint model for several Ngrams, 
but those efforts are outside the scope of this illustration.
Naturally, the whole analysis rests upon a strong assumption: 
that the change-point parameter underlying the sequence 
of battle-deaths and the (potential) change-point parameter 
underlying the Ngram are somehow the same parameter. 
We thus assume that changes in the battle death distribution 
and in the `anti-war' distribution are two different 
manifestations of the same underlying process.  

We model the `anti-war' Ngram with a simple normal model
with an autoregressive correlation structure of order 1.
We allow the change-point to influence both the expectation 
and variance parameters of the model, but it turns out that 
it is primarily the expectation that changes across 
the change-point (it increases). The correlation between 
consecutive years is high (0.80). From Figure \ref{figure:data} 
it is clear that there are at least one very clear 
change-point in the sequence of usage rates: 
from 1823 to 1914 `anti-war' is hardly used at all 
(dashed line in the figure), and then the use increases. 
This increase might reflect a genuine increase in usage, 
or simply that the Google Books corpus is less complete 
for older texts. At any rate, we will assume that the 
change-point around 1914 (from no use to some use) 
is not the one we are interested in, but rather that 
the change in norms we are searching for must be
reflected in a potential later change-point 
(from some use to more use). We will therefore only 
use the Ngram data for the years after 1914; 
one must bear in mind that this entails that the 
Ngram can only influence the change-point inference 
for the latter part of the full sequence of war years. 

Using the autoregressive model and the method 
from \citet*{CHH18}, we obtain another log-likelihood 
profile $\ell_{N,\prof}(\tau)$ and also the full 
confidence curve based on the Ngram information 
(in blue in the left panel of Figure \ref{figure:ccBN}). 
This curve has a point estimate at 1962, but with 
considerable confidence for the change rather 
taking place in 1927, or in 1971.

In the fusion step, the most straightforward solution is simply 
to sum the two log-likelihood profiles, calculate the deviance,  
and run the simulations to find the distribution of the deviance 
at each potential change-point (in the way as described in \citet*{CHH18}). 
This raises the question on whether it is appropriate 
to treat the two sources of information equally, however. 
There could be good reasons to consider the battle death data 
to be more directly informative for $\tau$ than the `anti-war' data. 
These arguments invite a combined confidence log-likelihood of the form
\beqn
\hbox{FF}\colon \qquad 
\ell_{\fus}(\tau) 
   =w_B\ell_{B,\prof}(\tau)+ w_N \ell_{N,\prof}(\tau), 
\eeqn 
with weight factors $w_B,w_N$ reflecting the relative
importance attached to the N source of information 
compared to the B source. In the optimistic case 
where the two sources are seen to inform on 
the same parameter in equal measure, the weights
are set to 1. In most situations there might be 
no direct information available which could help one 
estimate these importance-or-relevance weights. 
In cases where these weights are set by the analyst,
the effect of the choice should be communicated 
clearly and openly. This is also related to 
difficult themes of what might be seen as 
a meaningful proxy for what. 

In the right panel of Figure \ref{figure:ccBN}
we display the confidence curve with balance parameters
$(1,0.2)$ (in light violet), along with the curve without 
down-weighting of the soft data, i.e.~weights $(1,1)$ 
(in dark violet). Both combined curves indicate
a point estimate of 1965; this was not the point estimate
in any of the two sources, but that year has
a high confidence in both. The combined curve with
equal balancing gives the appearance of higher precision 
than the light violet curve, but this might be misleading 
if we do not trust the `soft data' fully. Then we might 
prefer the combined curve with weights $(1,0.2)$, 
which is more similar to the original curve based
on the battle death information only.

We must emphasise that we do not recommend the automatic 
use of this subjective importance weighting for all
or most information combination settings. 
In usual settings, the `degree of informativeness' 
of each source is already sufficiently well represented 
by the likelihood component from that source. However, 
down-weighting of softer data can be considered
in situations like the present one, with combination
of soft and hard data, where there might be
stark differences in quality or relevance between the sources.

In this application we have illustrated a situation 
where one information source was considered to be of 
higher quality and relevance than the other; 
a combination of hard and soft data. Note that such combination 
attempts often require the users to make strong assumptions, 
for example that very different sources inform on 
the exact same parameter. Low-quality, large-data sources 
are expected to play an increasingly important role 
in statistics in years to come (especially via scraping 
of the internet). The combination of such data sources 
with more high-quality sources raises various issues, 
and we will end with a note of caution. In a best case scenario, 
the analyst manages to benefit from a large, 
low-quality source and can obtain more precise statements 
than those from the smaller, high-quality sources alone. 
In the worst case scenario, the analyst is contaminating 
good data with irrelevant noise, and does not learn 
anything of value. 

\section{Illustration of Cox--Reid in the II step:
  the Neyman--Scott problem}
\label{section:appendixA}

The Neyman--Scott problem is an extreme example 
of such a situation. It can be presented of as 
a meta-analysis problem. We have a large number $k$
of studies, but each study has only two observations 
(so $n_1=\dots=n_k=n=2$). From each source $j$ 
we observe $y_{i,j} \sim \N(\mu_j, \sigma^2)$ 
where $i=1,2$ and $j=1,\dots,k$. Each sources has a specific 
mean parameter, but the variance, which is the parameter of main 
interest, is common across sources. This problem is popular in the 
literature concerning corrections to the profile likelihood 
(see e.g.~\citet[Ch.~7]{CLP}), since there exists a 
simple and exact solution, which serves as a gold standard 
against which to compare different corrections. 
We will compare this gold standard solution, 
as found in \citet[Ch.~4]{CLP}, 
against the `standard' II-CC-FF solution and the corrected 
II-CC-FF solution using the simple Cox--Reid correction.

\begin{figure}[h] 
\begin{center}
\includegraphics[scale=0.4]{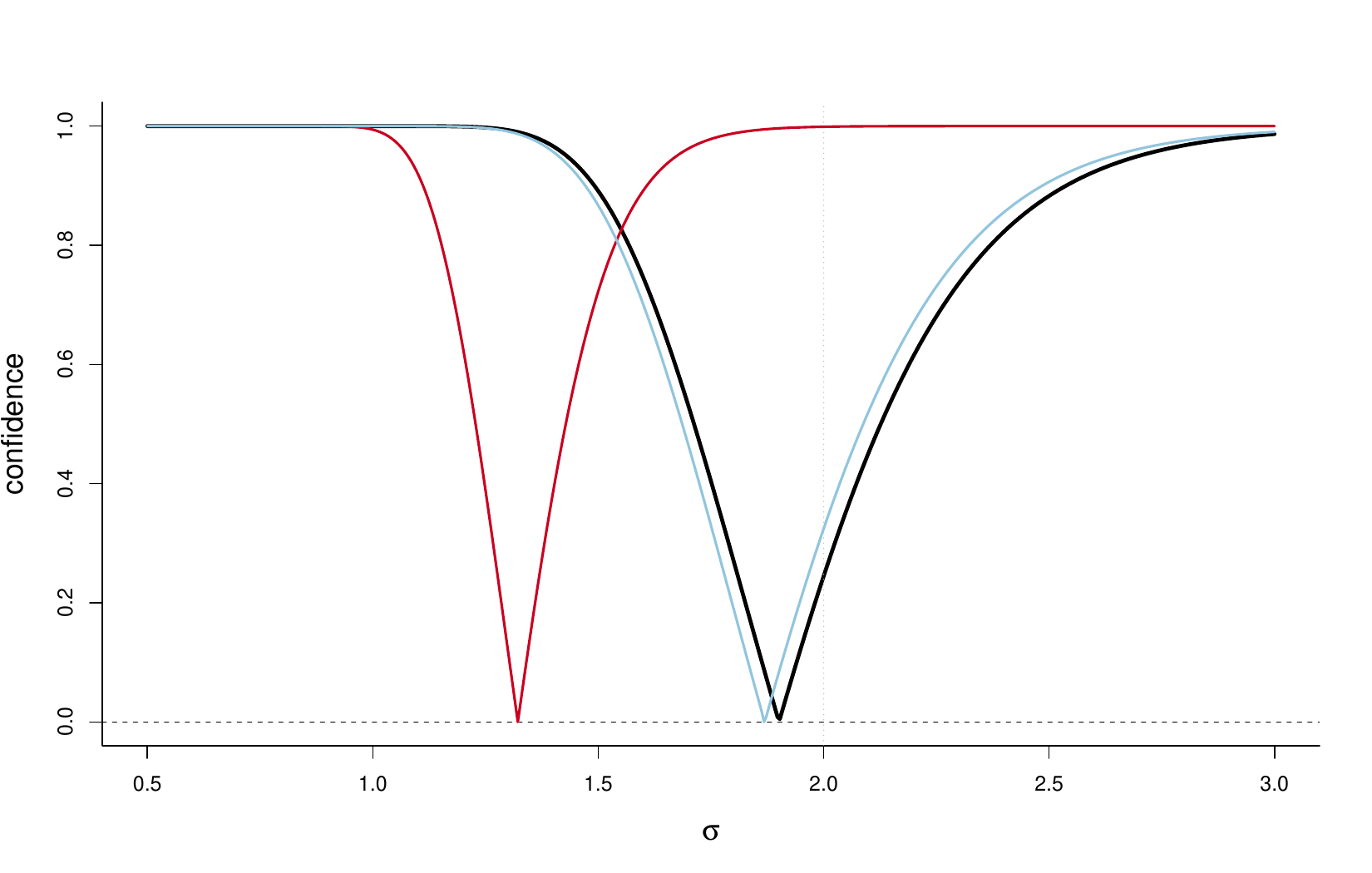}
\end{center}
\caption{Neyman--Scott example with $k=20$ sources, 
the parameter of main interest $\sigma=2$, 
and the source specific means drawn from a uniform 
between $-3$ and 3. The exact confidence curve 
in black, the standard uncorrected II-CC-FF solution in red, 
and the corrected II-CC-FF solution in blue.}
\label{figure:neymanScott} 
\end{figure}

The pivot $\hatt \sigma^2/ \sigma^2$ can be seen to have a 
$\chi_{k}^2/(2k)$ distribution, which gives the exact CD,
$$C_{\gold}(\sigma) = 1 - \Gamma_k(2k \hatt \sigma^2/ \sigma^2),$$
where $\Gamma_k(\cdot)$ is the c.d.f.~of the $\chi_{k}^2$ distribution 
and 
$\hatt \sigma^2 = \sumjk S_j^2/(2k)   =  \sumjk \half (y_{1,j}-y_{2,j})^2/(2k)$ 
is the ML estimate (we note that this is a famous case 
where the ML estimator is inconsistent). 
The gold standard CD may be turned into a confidence curve 
using \eqref{eq:cc} and is displayed in black in 
Figure \ref{figure:neymanScott}. For the sake of comparisons, 
we can write out the confidence likelihood implied by 
this CD,
\beq
\ell_{\con,\gold}(\sigma) = - k\log\sigma - \half (1/\sigma^2) 
   \sumjk S_j^2. 
\label{eq:llopt}
\eeq

With the standard II-CC-FF solution we start with the II step 
where we deal with each source separately: we profile out 
$\mu_j$ and get 
$\ell_{\prof,j}(\sigma) = -2\log\sigma - \half (1/\sigma^2) S_j^2$.
All the sources inform on exactly the same focus parameter 
and we can just sum the log-likelihood contributions in the fusion step, 
\beq
\ell_{\prof}(\sigma) = -2 k \log\sigma 
   - \half (1/\sigma^2) \sumjk S_j^2. 
\label{eq:llp}
\eeq
Comparing this to the confidence log-likelihood for the gold standard 
in \eqref{eq:llopt} we see that they differ by an extra `2' 
in the first term, which causes the inconsistency of the ML estimator. 
We may nevertheless construct our confidence curve 
in the general II-CC-FF manner,
$\cc^{*}_1(\sigma,\data) = \Gamma_1\bigl(2\{\ell_{\prof}(\hatt\sigma) 
   - \ell_{\prof}(\sigma)\}\bigr)$.

For this model, the simple Cox--Reid correction term 
for each source is $\log\sigma$. The corrected profile log-likelihood 
for each source is then $\ell_{\prof,j}(\sigma)+\log\sigma$, 
and for the full data we obtain a corrected profile log-likelihood 
identical to \eqref{eq:llopt}. 
Following the II-CC-FF recipe we construct the confidence curve 
with the Wilks approximation, 
$\cc^{*}_2(\sigma,\data) = \Gamma_1\{2(\ell_{\con,\gold}(\hatt \sigma) 
   - \ell_{\con,\gold}(\sigma)) \}$.

Figure \ref{figure:neymanScott} gives the three confidence curves 
in a specific example with $k=20$ sources. The standard II-CC-FF 
solution in red is clearly far from the exact black curve. 
With $k$ increasing, it converges to the wrong value, $\sigma/\sqrt{2}$. 
The blue curve, on the other hand, corresponding to the II-CC-FF 
solution with Cox--Reid correction, is close to the exact curve, 
even though it is constructed using the Wilks approximation. 
When $k$ increases, for instance to 50, 
the blue and black curves are virtually identical.  

\end{document}